\documentclass[a4paper,prl,reprint, twocolumn,superscriptaddress]{revtex4}

\usepackage{ulem}
\usepackage{amssymb}
\usepackage{amsmath}
\usepackage{epsfig}
\usepackage{color}
\usepackage{graphics, graphicx}
\usepackage{bbold}
\usepackage{psfrag}
\usepackage{mathcomp}
\usepackage{amsmath}
\usepackage{amssymb}
\usepackage{mathrsfs}
\usepackage{subfigure}
\usepackage{verbatim}
\usepackage{xcolor}

\usepackage[colorlinks,citecolor=blue,urlcolor=blue]{hyperref}

\usepackage{mathrsfs}

\begin{document}

\date{\today}
\title{Robust quantum control for higher order coupling term in trapped ions}
\author{Jing-Bo Wang}
\email{wangjingbo02@baidu.com}
\affiliation{Institute for Quantum Computing, Baidu Research, Beijing 100193, China}

\begin{abstract}
Trapped ion hardware has made significant progress recently and is now one of the leading platforms for quantum computing. To construct two-qubit gates in trapped ions, experimental manipulation approaches for ion chains are becoming increasingly prevalent. Given the restricted control technology, how implementing high-fidelity quantum gate operations is crucial. Many works in current pulse design optimization focus on ion-phonon and effective ion-ion coupling while ignoring the higher-order expansion impacts of these two terms brought on by experiment defects. This paper proposed a novel robust quantum control optimization method in trapped ions. By introducing the higher-order terms caused by the error into the optimization cost function, we generated an extremely robust Mølmer–Sørensen gate with infidelity below $10^{-3}$ under drift noise range $\pm 10$ kHz and time noise range $\pm 0.02$. Our work reveals the vital role of higher-order coupling terms in trapped ion pulse control optimization, especially the higher ion-ion coupling order, and provides a robust optimization scheme for realizing more efficient entangled states in trapped ion platforms.
\end{abstract}

\maketitle

\section{I. Introduction}
In recent years, the trapped-ion quantum computing platform has developed rapidly~\cite{bruzewicz2019trapped,mehta2020integrated,niffenegger2020integrated,grzesiak2020efficient,linke2017experimental}, and the performance of trapped-ion chips has shown an explosive growth pattern. The best state preparation fidelity, measurement fidelity, and gate fidelity have been accomplished on trapped-ion platforms~\cite{srinivas2021high,ballance2016high,gaebler2016high}. The rapid growth trend shows trapped-ion platforms are one of the leading candidates for building the noisy intermediate-scale quantum devices ~\cite{preskill2018quantum,zhang2017observation}, and has demonstrated promising findings in quantum chemistry and small-scale molecular simulations~\cite{kawashima2021optimizing, hempel2018quantum,nam2020ground}. Furthermore, fault-tolerant quantum computing requires gate infidelity below $10^{-2}\sim 10^{-4}$ is also available in trapped-ion platforms~\cite{campbell2017roads,linke2017fault,bermudez2017assessing,benhelm2008towards}. 

In trapped-ion hardware, ions can be confined in a trap known as the Paul trap, which consists of radio-frequency and direct current electrodes~\cite{paul1990electromagnetic}. Many advanced construct technologies, such as two-dimensional Penning traps, and quantum charge-coupled devices, have been developed~\cite{brown1986geonium,britton2012engineered,kielpinski2002architecture,bowler2012coherent,pino2021demonstration}. To complete the specific quantum computing algorithms on trapped-ion platforms, we need to perform appropriate operations on the ion. Take $^{171}$Yb$^+$ ion as an example, qubit states $\{|0\rangle, |1\rangle\}$ are encoded into ion's hyperfine internal states such as $\{|F=0, m_F=0\rangle, |F=1, m_F=0\rangle\}$. For single-qubit operations, we can conveniently flip ions between $|0\rangle$ and $|1\rangle$ states with two bichromatic Raman lasers.  For multi-qubit operations, through the phonon mode caused by the collective vibration of ion chains, we can transform quantum information between different ions and create the native multi-qubit gates.  

With the increased qubit number in trapped-ion platforms and complex quantum operations required for quantum advantage, higher requirements are put forward for the experimental setup and pulse optimal control methods~\cite{choi2014optimal,myerson2008high,srinivas2021high,harty2014high}. With ion qubits trapped in the vacuum cavity, the phonon vibration frequency used to transfer information will change with the ion's position, environment setup, and laser pumping~\cite{wu2018noise}. The natural pulse parameters are also significantly deformed due to the laser generator's accuracy limitations. How to effectively design pulse control optimization with various environmental noises are indispensable for future trapped-ion quantum computers. Many studies focus on applying various pulse sequence strategies to construct a robust high-fidelity Mølmer-Sørensen (MS) gate between trapped ion qubits~\cite{hayes2012coherent,haddadfarshi2016high,manovitz2017fast,shapira2018robust,webb2018resilient,zarantonello2019robust}, since all these parameters in the experimental environment would significantly affect the operation accuracy. Remarkably, two recent works have realized a frequency noise robust MS gate using different pulse modulation methods~\cite{ruzic2022frequency, jia2022angle}.  For better performance of MS gates in trapped-ion platforms, firstly, we need to enhance the trapped equipment's accuracy to solve the environmental interference problem. More importantly, we must design trapped-ion robust optimization pulse control methods insensitive to many environmental noises.\\

This paper demonstrates a simple and efficient pulse control optimization method to realize a robust MS gate against expansive frequency drift and time noise. Additionally, our calculation reveals the importance of higher ion-ion coupling order response. In Sec. II, we briefly introduce the basic quantum control theory and some optimization methods of trapped ions. Then, in Sec. III, we carefully analyze the effect of higher-order ion-phonon and ion-ion coupling terms on MS gate fidelity and define a suitable cost function. In Sec. IV, we numerically calculate the MS gate performance, residual ion-phonon coupling, and effective ion-ion coupling at a wide noise range under our optimizing quantum control method. Furthermore, we give the corresponding experimental parameters in the existing laboratory platforms. Finally, in Sec. V, we summarize our robust control scheme and expect our strategy to provide better solutions for trapped-ion MS gates in future experiments.

\section{II. Entangled gates and fidelity in trapped ions}
In universal quantum computing circuit models, complex quantum computing algorithms can decompose into a series of quantum gates, including single-qubit, two-qubit, or multi-qubit gates~\cite{nielsen2002quantum}. Different hardware platforms, for instance, superconductors or trapped ions, may prefer various control methods. Currently, we adopt laser control methods to realize the native quantum gate in trapped-ion quantum computing~\cite{srinivas2021high,ballance2016high}. The native quantum gate refers to the easily implemented quantum operation on such a hardware device; for trapped ions, the native quantum gates include rotating gates $\{R_x, R_y, R_z\}$ and MS gate $\{e^{-i\frac{\pi}{4}\sigma_x\otimes\sigma_x}\}$~\cite{sorensen1999quantum}. We can use bichromatic Raman laser beams to form a Rabi frequency and implement a single-qubit gate. Several experiments use amplitude modulation methods, which modulate Rabi frequency amplitudes $|\vec\Omega(t)|$~\cite{zhu2006arbitrary,roos2008ion}. Furthermore, we can use multi-individually addressed laser beams to implement multi-qubit gate operations in trapped ions. The qubit decouples from multiple phonon modes in the ion chains, and related ion qubits are practical couples~\cite{steane2014pulsed,leung2018robust,leung2018entangling,lu2019global,landsman2019two}.\\

Other works may focus on phase modulation by just adjusting the phase term $\phi(t)$ of the Rabi frequency. We can suppress the dominant noise source analytically and numerically with separate phase slices~\cite{milne2020phase,leung2018robust}. Moreover, we can decouple separated ions and phonon modes while connecting multi-ion qubits simultaneously. Nevertheless, the exponential growth of the pulse number is unacceptable since we cannot fabricate such a device when the ion number is more extensive, and the operation requirement has exceeded the capacity limit of the laser device at this stage. A more efficient way is adjusting the amplitude and phase of the Rabi frequency simultaneously. However, recent experiments and optimization works give a solution that suppresses noise only within a narrow range of drift noise parameters~\cite{bentley2020numeric}.\\

Here, we briefly review the implementation of two-qubit MS gates in trapped ions from the Hamiltonian perspective. Based on this, we deduce the fidelity of the MS gate and the dominant coupling terms. For convenience, we set $\hbar = 1 $, the effective Hamiltonian by implementing a global laser and an individual laser is~\cite{green2015phase} 

\begin{align}
H(t)=i\sum_j \Omega_j(t)\sigma_x ^j \sum_k \eta_{j, k}(a_k e^{i(\mu-v_k) t}e^{i\phi_j(t)}-h.c),
\end{align}
where we define the efficient Rabi frequency amplitude as $\Omega_j(t)$ , and relative Rabi frequency phase as $\phi_j(t)$, $\sigma_x^j$ represent the internal state of $j_{th}$ ion. The summation of $j$ reveals that the phonon modes connect all the laser-implemented ions, and $a_k$ is the $k_{th}$ phonon annihilation operator. In laser-driven trapped ion quantum platforms, the Lamb-Dicke parameter $\eta_{j,k}=\Delta K b_{j,k} \sqrt{\frac{1}{2Mv_k}}$ is small for all $\{j,k\}$ indexes. $\Delta K$ is the wave vector difference and $M$ is ion's mass, $b_{j,k}$ is the micro-motion of $j_{th}$ ion response to $k_{th}$ phonon mode, which can calculate by participation perturbation theory~\cite{james1998quantum}. We can set the laser detuning as $\mu$, which has a discrepancy with phonon frequency $v_k$. In $\eta_{j,k}\ll 1$ regime, by using Magnus expansion, the evolution of the effective Hamiltonian is 

\begin{align}
U(\tau)=&\exp \{- i\int_0^\tau H(t)dt-\frac{1}{2}\int_0^\tau \int_0^{t_1} [H(t_1),H(t_2)] dt_1 dt_2\notag\\
&-\frac{i}{6}\int_0^\tau \int_0^{t_1}\int_0^{t_2} [H(t_1),[H(t_2),H(t_3)]]dt_1 dt_2 dt_3 + \cdots\},
\end{align}
where $\tau$ is the quantum gate duration time. In the Raman-driven model, the process occurs in the Lamb-Dicke regime, and we can ignore the higher 2-order expansion terms. For $-i\int_0^\tau H(t) dt$, define $B_k=(\mu-v_k)$, which is usually called sideband frequency. We derive the coefficient for ion-phonon coupling term $a_k\sigma_x^j$

\begin{align}
\beta_{j,k}(\tau) = & \int_0^\tau \Omega_j(t)e^{i\phi_j(t)} \eta_{j,k} e^{iB_k t} dt.
\label{eq:beta}
\end{align}
While for the second expansion term $[H(t_1),H(t_2)]$, the expansion is
\begin{align}
[H(t_1),H(t_2)] = &-2i \sum_{j\neq j'}\sum_{k}\Omega_j(t_1)\Omega_{j'}(t_2) \eta_{j,k} \eta_{j',k}\times\notag\\
& \sin\big[B_k(t_1-t_2)+\phi_j(t_1)-\phi_{j'}(t_2)\big]\sigma_x ^j \sigma_x ^{j'}.
\end{align}
Then we can define integral 

\begin{align}
\theta_{j,j'} (\tau)=& \sum_{k}\int_0^\tau \int _0^{t_1}dt_1 dt_2\Omega_j(t_1)\Omega_{j'}(t_2) \eta_{j,k} \eta_{j',k}\times\notag\\
& \sin\big[B_k(t_1-t_2)+\phi_j(t_1)-\phi_{j'}(t_2)\big],
\label{eq:theta}
\end{align}
which represents the effective ion-ion coupling strength between the internal states of ion $j$ and $j'$, then the total evolution matrix $U(\tau)$ can be rewritten as

\begin{align}
U(\tau)=\exp \left[\sum_{j, k}\big( \beta_{j, k}(\tau) a_k \sigma_{\mathrm{x}}^{j} - h.c\big) +\mathrm{i} \sum_{j\neq j'} \theta_{j, j^{\prime}}(\tau) \sigma_{\mathrm{x}}^{j} \sigma_{\mathrm{x}}^{j^{\prime}}\right].
\end{align}
If $\sum_{j\neq j'}\theta_{j,j'}(\tau)=\pi/4$, and $\beta_{j, k}(\tau)=0$ for all phonon $k$ coupled with ion $j$, then two ions reach a maximum entanglement with no residual ion-phonon coupling. The MS gate fidelity can been written as $F=\big|\frac{\text{Tr}(U_{g}^\dagger U(\tau))}{D} \big|^2$, where D is the dimension of unitary $U_{g}^\dagger U(\tau)$, $U_g$ is the goal unitary. The total unitary can be separated into two parts $U(\tau)=U_1(\tau)U_2(\tau)$, which are denoted by the notation
\begin{align}
U_1(\tau)=&\exp\Big(i\sum_{j\neq j'} \theta_{j, j^{\prime}}(\tau) \sigma_{\mathrm{x}}^{j}\sigma_{\mathrm{x}}^{j^{\prime}}\Big),\\
U_2(\tau)=&\exp\Big(\sum_{j, k}\big( \beta_{j, k}(\tau) a_k \sigma_{\mathrm{x}}^{j} - h.c\big)\Big).
\end{align}
Here $U_1(\tau)$ is the ion-ion coupling unitary, and $U_2(\tau)$ is the ion-phonon coupling unitary. While in the thermal product formula
\begin{align}
\rho_k=&\sum_{n_k=0}^\infty |n_k\rangle\langle n_k| (1-e^{-\hbar v_k/k_B T})e^{-n_k\hbar v_k/k_B T},\\
\rho=&\rho_1\otimes\cdots\otimes\rho_N,
\end{align}
where $\rho_k$ is the $k_{th}$ phonon density matrix and $\rho$ is the total thermal states density matrix, $|n_k\rangle$ is $k_{th}$ phonon occupancy state, here we considering a finite temperature $T$ and define Boltzmann constant as $k_B$. By tracing over the phonon density matrix, the residual ion-phonon coupling influence on MS gate fidelity could write as
\begin{align}
\langle U_2(\tau)\rangle_\rho=\prod_k e^{-\sum_j|\beta_{j, k}(\tau)|^2(\overline {n}_k+1/2)},
\end{align}
the average phonon occupation number can be written as $\overline {n}_k=\frac{e^{-\hbar v_k/k_B T}}{1-e^{-\hbar v_k/k_B T}}$~\cite{roos2008ion}, by combining the trace of $U_g^\dagger U_1(\tau)$, the final fidelity of MS gate in trapped ions is~\cite{bentley2020numeric}
\begin{align}
F = |\cos(\sum_{j\neq j'}\theta_{j,j'}(\tau)-\pi/4)|\prod_k e^{-\sum_{j}|\beta_{j,k}(\tau)|^2(\overline {n}_k+1/2)}.
\label{eq:fidelity}
\end{align}
For a high-fidelity MS gate, we need to ensure that the ion's internal states will decouple with phonons within a gate period after the laser pulse sequences. In order to avoid quantum information loss during quantum control operations, a major purpose for suitable control optimization is to make $\beta_{j,k}(\tau)\to 0$ for all ion $j$ and phonon $k$ and make the ion-ion coupling strength $\sum_{j\neq j^\prime}\theta_{j, j^\prime}(\tau) \to \pi/4$. All the effort to reduce $(\sum_{j\neq j^\prime}\theta_{j,j^\prime}(\tau)-\pi/4)$ and $|\beta_{j,k}(\tau)|^2$ will increase the gate fidelity.\\

\section{III. Robust optimization methods in trapped ion quantum control}
As we demonstrated above, for MS gates in trapped ions, we can design a pulse sequence that completely decouples ion phonon coupling strength $\beta_{j,k}(\tau)\to 0$. Furthermore, when efficient ion-ion coupling strength $\sum_{j\neq j^\prime}\theta_{j, j^\prime}(\tau) \to \pi/4$, we can maximumly entangle two implemented ions. However, such quantum gates usually endure the drift of sideband frequency $B_k$. If the drift noise change like $B_k\to B_k+\delta$ where $\delta \ll 1/\tau $. For ion-phonon coupling, $\beta_{j, k}(\tau)$ will suffer the following error
\begin{align}
\Delta\beta_{j,k}(\tau)\approx&\frac{\partial \beta_{j,k}(\tau)}{\partial B_k} \delta+ \mathcal{O}(\delta^2)=\widetilde \beta_{j,k}(\tau)\delta + \mathcal{O}(\delta^2),
\end{align}
where we define
\begin{align}
\widetilde \beta_{j,k}(\tau)=\frac{\partial \beta_{j,k}(\tau)}{\partial B_k}=i\int_0^\tau \Omega_j(t)e^{i\phi_j(t)} \eta_{j,k} e^{iB_k t}t dt.
\label{eq:betaOrder1}
\end{align}\\

For quantum control in trapped ions, firstly, we need to make $\beta_{j, k}(\tau) \to 0$. In the recent experiment simulation, amplitude and phase hybrid control have been used in trapped ion experiments to decouple the complicated phonon mode in long ion chains and have realized the maximum entangled state~\cite{bentley2020numeric}. For laser devices, the waveform envelope can arrange using the pulse sequence technology, where the equant slice duration time $\tau_s=\tau/L$, $L$ is the total sequence number. In the amplitude and phase hybrid modulating method, if we set phase/amplitude term as $\phi_j (t)=\{\phi_{j,1}, \phi_{j,2}, \cdots, \phi_{j, L}\}$ and $\Omega_j(t)=\{\Omega_{j,1}, \Omega_{j,2}, \cdots, \Omega_{j, L}\}$. By defining $S_{j,k}^l=\eta_{j,k}\Omega_{j,l}e^{i\phi_{j,l}}$, the ion-phonon coupling term $\beta_{j, k}(\tau) $ can be written as:
\begin{align}
\beta_{j, k}(\tau) = &\eta_{j, k} \sum_{l=1}^L\int_{(l-1)\tau_s}^{l \tau_s}dt  \Omega_{j, l}  e^{i\phi_{j, l}} e^{iB_k t}  \notag\\
=&-i\sum_{l=1}^L S_{j,k}^l\frac{e^{iB_kl\tau_s}}{B_k}(1-e^{-iB_k\tau_s}).
\end{align}

Since we focus on an arbitrary ion chain, for example, a linear 1-dimensional chain, the phonon transverse mode number $M$ has a relation $M=N$ with the ion number $N$, and the phonon frequency has a relation $v_1:v_2:v_3:v_4 \cdots \approx \sqrt {4.6}:\sqrt {6.1}:\sqrt {7.5}:\sqrt {8.5} \cdots $~\cite{james1998quantum}. So it is not possible to make $1-e^{-iB_k\tau_s}$ equal 0 for all phonon modes. The best effort we can make is  $B_1\tau_s=2\pi$, then the largest phonon-ion coupling term equals 0 with any pulse shape, while other modes remain residual. Because the $1-e^{-iB_k\tau_s}$ term is a constant not equal to 0 for long ion chain, then we need to adjust the $S_{j,k}^l$ to make sure $\sum_{l=1}^L S_{j,k}^l\frac{e^{iB_k l\tau_s}}{B_k} \to 0$. Moreover, although drift noise $\delta$ is relatively small compared to detuning frequency, there is no theoretical guarantee that the previous partial derivative $\widetilde \beta_{j,k}(\tau)$ is also small. For robust quantum control, we need to make $\Delta\beta_{j,k}(\tau) \to 0$, which means $\widetilde \beta_{j,k}(\tau) \to 0$ for all $(j, k)$ indexes.\\

While for the effective ion-ion coupling $\theta_{j,j^\prime}(\tau)$, which also plays an important role in MS gate fidelity as shown in Eq. (\ref{eq:fidelity}), the drift noise cause entangled phase error
\begin{align}
\Delta\theta_{j,j'} (\tau)\approx \frac{\partial \theta_{j,j^\prime}(\tau)}{\partial B_k} \delta+\mathcal{O}(\delta^2)=\widetilde \theta_{j,j'} (\tau) \delta+\mathcal{O}(\delta^2),
\end{align}
where the first-order coefficient is
\begin{align}
\widetilde \theta_{j,j'} (\tau)=&\sum_{k}\int_0^\tau \int _0^{t_1}dt_1 dt_2\Omega_j(t_1)\Omega_{j'}(t_2) \eta_{j,k} \eta_{j',k} \times\notag\\
&\cos\big[B_k(t_1-t_2)+\phi_j(t_1)-\phi_{j'}(t_2)\big] (t_1-t_2).
\label{eq:thetaOrder1}
\end{align}
If $\widetilde \theta_{j,j'} (\tau)$ is very large, even small drift noise $\delta$ will make a substantial change to the gate performance. \\

It is inappropriate to directly optimize the effect of noise on MS gate fidelity in Eq. (\ref{eq:fidelity}) for its complex formula. As the analysis above, we notice that all efforts to make Eq. (\ref{eq:beta}) (\ref{eq:betaOrder1}) (\ref{eq:thetaOrder1}) approach 0 and summation of Eq (\ref{eq:theta}) approach $\frac{\pi}{4}$ will contribute to MS gate performance. Since $\{\beta_{j,k}(\tau), \widetilde \beta_{j,k}(\tau)\}$ have linear relation with $\Omega_j(t)$ integral, and $\{\theta_{j,j^\prime}(\tau), \widetilde \theta_{j,j^\prime}(\tau)\}$ have quadratic relation with $\{\Omega_j(t),\Omega_{j^\prime}(t)\}$ integral. For our quantum control optimization method, it is suitable to define a cost function
\begin{align}
\mathcal{C}=&\sum_{j,k}\Big(|\beta_{j,k}(\tau)|^2+|\widetilde \beta_{j,k}(\tau)|^2\Big)\notag\\
&+\Big(|\sum_{j\neq j^\prime}\theta_{j,j^\prime}(\tau)-\pi/4|+\sum_{j\neq j^\prime}|\widetilde \theta_{j,j^\prime} (\tau)|\Big),
\label{Costfun}
\end{align}
by optimizing $\mathcal{C}$, we can find the pulse sequence which ensure $\{\beta_{j,k}(\tau),\theta_{j,j^\prime}(\tau)\}$ insensitive to drift noise. Then, we can calculate a full MS gate fidelity by using Eq (\ref{eq:fidelity}) under the drifting noise $\delta$ or time noise $\tau \to \widetilde \tau$. In a trapped ion chain with $N$ ions, we notice that the minimum parameter number required to optimize Eq. (\ref{Costfun}) is $4N +2 $ for MS gates, so a linearly growing parameter space can achieve. Here we choose $5N$ for our robust optimization.\\

Our optimization model includes two new terms, one representing higher-order ion-phonon coupling and the other representing efficient higher-order ion-ion coupling. For an appropriate cost function $\mathcal{C}$. The optimized pulse control parameters achieve robustness to ambient noise. In the next section, we provide a detailed numerical demonstration of our scheme's impact on an achievable experimental parameter.

\section{IV. numerical results of robust optimization}
\subsection{A. Experimental simulation setup}
In trapped ion quantum control, the amplitude, phase, and frequency of the effective Rabi frequency are all adjustable parameters. Due to the monotonic growth of coupling strength in pure amplitude modulation and easily phase-space trajectory change in phase modulation. For our numerical optimization scheme, we adopt amplitude $\Omega(t)$ and phase $\phi(t)$ simultaneously while fixing the frequency. The hybrid modulating amplitude and phase help phonons quickly transform quantum information between two ions. To better achieve the optimized effect, for the amplitude sequence, we adopt a symmetrical design and set the amplitude as $\Omega_j(t) = \Omega_j(\tau-t)$, while for the phase sequence, we adopt an anti-symmetric design $\phi_j(t) = -\phi_j(\tau-t)$. \\

To ensure that our simulation results match the experimental requirements, we set the trapped potential's axial frequency as $\omega_z=1.2$MHz and the vertical frequency as $\omega_x=\omega_y=3.6$MHz. We suppose 4-ytterbium ions ($^{171}\text{Yb}^+$) are trapped by the potential and form a 1-dimensional linear array under the repulsive Coulomb interaction and negative trapped potential. Ions can be cooled down to $T=10^{-6}$ Kelvin by several cycles of initial sideband cooling. In such an experiment setting, we adopt transverse phonon modes as bus qubits between two ions. We can easily calculate the four transverse phonons mode frequency $v_k =\{2.5\text{MHz},2.96\text{MHz},3.29\text{MHz},3.5\text{MHz}\}$ as well as the related Lamb-Dick parameter $\eta_{j,k}$. For convenience, during our numerical optimization, we set laser detuning as $\mu=3.15$MHz and the gate duration time as $\tau = 100\mu \text{s}$. To match the power of the laser device, we set the maximum Rabi frequency amplitude $|\Omega_{\max}(t)|=2$MHz.

\begin{figure}[tbp!]
 \centering
 \includegraphics[width=8.5cm,height=6.5cm]{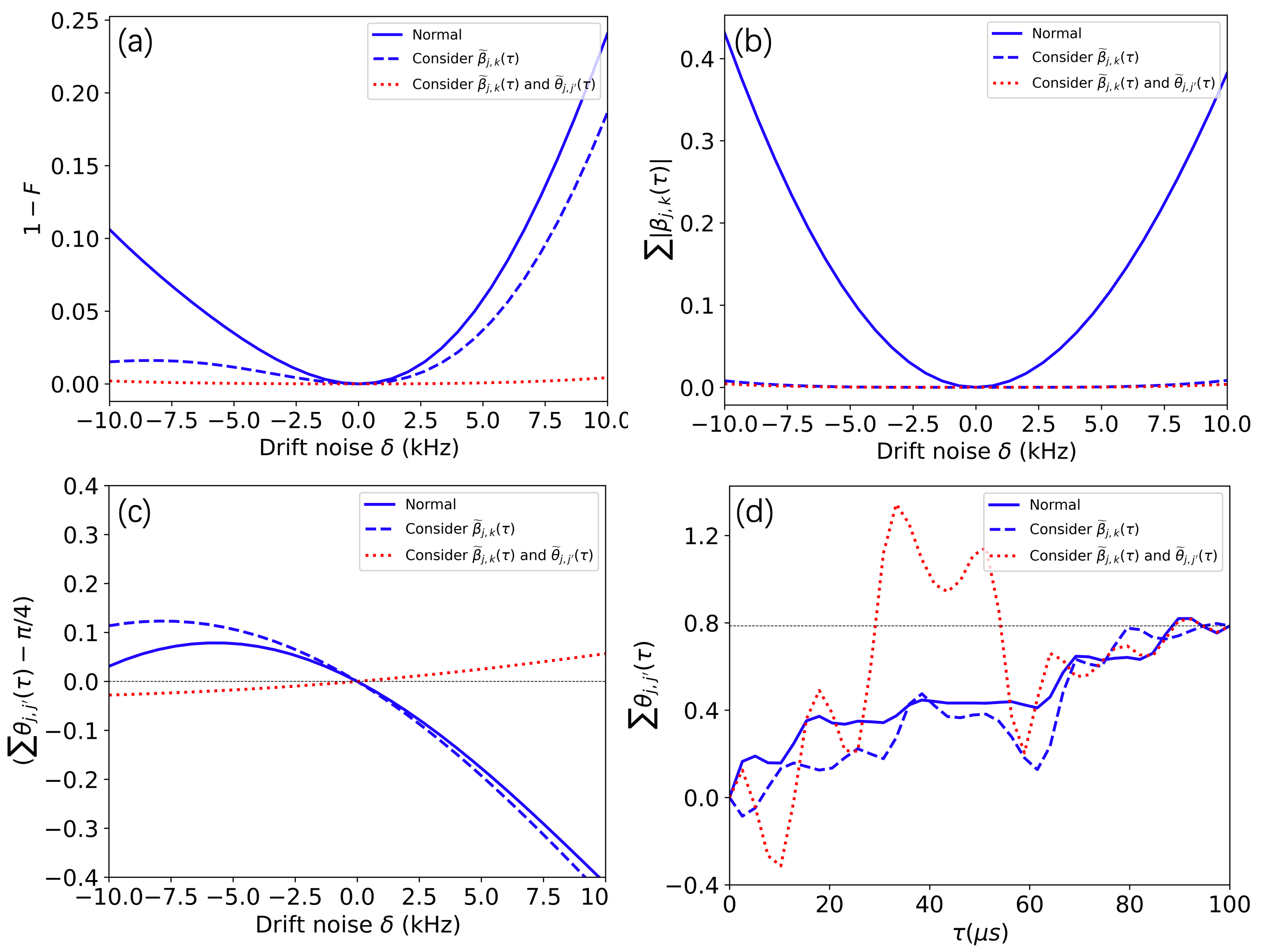}
 \caption{The robustness effect for our control optimization scheme on four ion chains. We apply the pulse to ion (1, 3) and calculate three optimization methods. We assume a drift noise range to $\pm 10$kHz and set gate duration time as 100$\mu s$. (a) Shows gate infidelity according to drift noise. (b) The total coupling strength of ions-phonon term with drift noise. (c) The effective ion-ion coupling change with drift noise. (d) The entangled coupling strength of two ions with duration time at $\delta=0$. It shows that higher-order terms play a vital role in optimization pulse design. When considering higher order terms, we notice two orders of magnitude in infidelity than the normal optimization. }
 \label{fig:Infidelity}
\end{figure}

\begin{figure*}[tbp!]
  \centering
  \includegraphics[width=16cm,height=4.5cm]{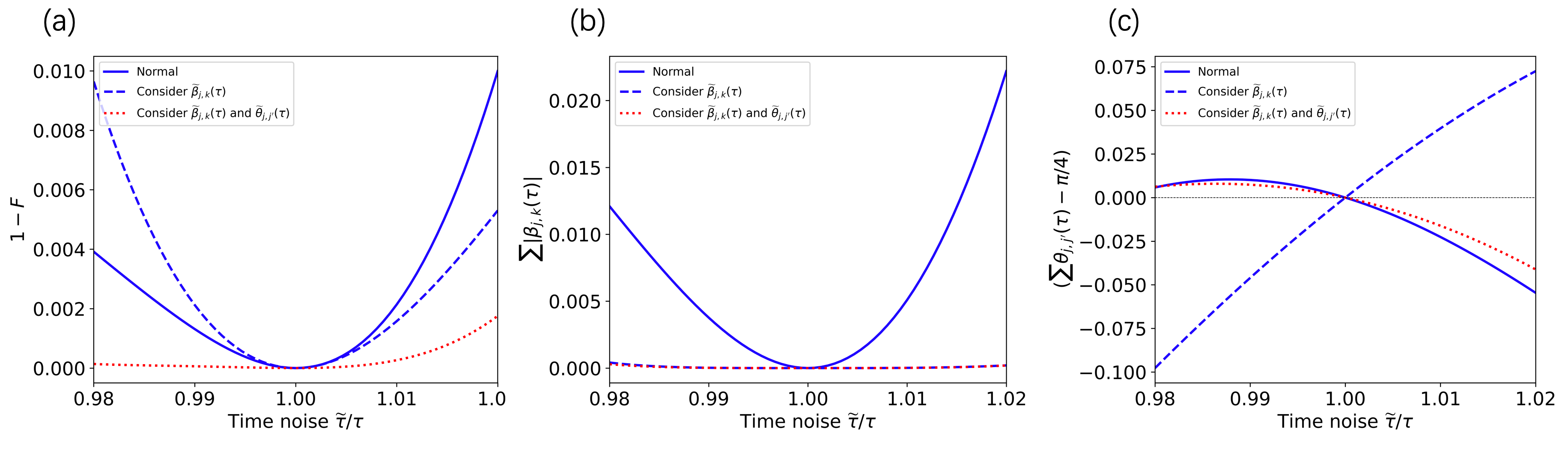}
  \caption{The effect of robust control optimization scheme to time noise, all the numerical simulation parameters are the same as in fig~\ref{fig:Infidelity}. (a) Shows gate infidelity according to time noise. (b) The total coupling strength of ion-phonon term with time noise. (c) The effective ion-ion coupling with time noise. While considering higher order terms, we still notice two orders of magnitude in infidelity than the normal optimization for time noise.}
  \label{fig:timenoise}
\end{figure*}

\subsection{B. Robust for drift noise}
In the numerical experimental environment, although we have set the laser frequency, the laser frequency may drift during the gate duration time. Furthermore, the external electromagnetic field will also cause a subtle change in the trapped potential well; such a disturbance will cause the phonon frequency to drift. All these lead to a variation of $B_k$ during the gate operation, generally called drift noise.

First, through the cost function defined in Eq~(\ref{Costfun}), we numerically calculated the corresponding pulse sequence using the SLSQP optimization method. Then using Eq~(\ref{eq:fidelity}), we calculated the fidelity of the optimized pulse sequence acting on two individual addressing ions, which can form MS gates in trapped ions. We also calculated the pulse sequence with the same optimization method for the other two cases and compared the robustness of the MS gate infidelity to drift noise for three cases. In a normal solution, we only consider that $\beta_{j,k}(\tau)$ and $(\sum_{j\neq j^\prime}\theta_{j,j^\prime}(\tau)-\pi/4)$ tend to 0 and define a normal cost function without $\widetilde \beta_{j,k}(\tau)$ and $\widetilde \theta_{j,j^\prime} (\tau)$ term in Eq~(\ref{Costfun}). In $\beta_{j,k}(\tau)$ robust solution, just consider the ion-phonon coupling term into first-order $\widetilde \beta_{j,k}(\tau)$ but ignore the higher ion-ion coupling order $\widetilde \theta_{j,j^\prime} (\tau)$ term. \\

We show the MS gate robustness to drift noise under the three pulse sequence optimizations. As we can see in Figure~\ref{fig:Infidelity} (a), after considering the higher order term of ion-phonon coupling $\widetilde \beta_{j,k}(\tau)$ and ion-ion coupling term $\widetilde \theta_{j,j^\prime} (\tau)$, the optimized pulse has the strongest robustness to drift noise. The MS gate infidelity is below $10^{-3}$ and has two orders of magnitude improvement compared with normal solutions. The ion-phonon residual coupling term can also be kept below 0.01 under 10kHz drift noise as shown in Figure~\ref{fig:Infidelity} (b), while for normal optimization method as shown by the solid blue line, the ion-phonon coupling is sensitive with drift noise. In Figure~\ref{fig:Infidelity} (c), we show the efficient ion-ion coupling strength change with drift noise in the optimized pulse sequence without considering the higher ion-ion coupling order term. With the drift noise increase, the coupling strength between ions will quickly deviate from the maximum entangled state, as shown by solid and dashed blue lines. The dashed red line shows that our optimization methods keep ion-ion coupling insensitive to drift noise. So, our methods can significantly improve the stability and practicability of the two-qubit gates in the trapped ion quantum computing hardware.

As shown in Figure~\ref{fig:Infidelity} (d), we calculated the variation trend of the efficient ion-ion coupling strength during the gate duration time. Our scheme makes the efficient ion-ion coupling strength exceed the strength required for maximum entanglement within the gate time, slowly fall back to the optimal entanglement strength, and finally plateaus. In the case of another two optimization schemes, the entanglement strength will slowly increase to the maximum entanglement strength.\\

\begin{figure*}[tbp!]
  \centering
  \includegraphics[width=16cm,height=4.5cm]{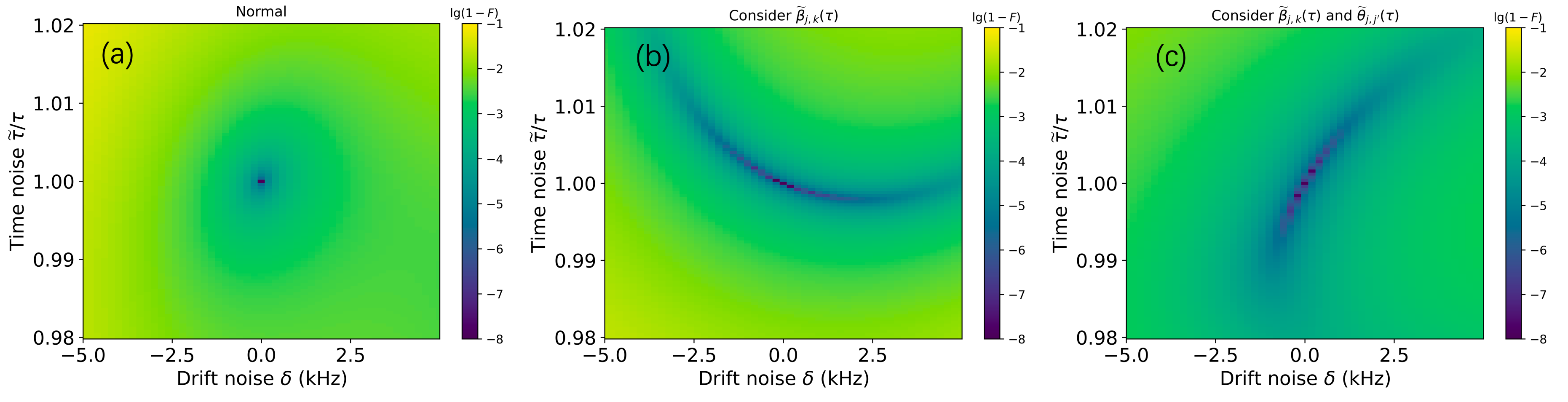}
  \caption{The robust performance of three optimization schemes under a combined drift and time noise, the drifting noise range to $\pm 5$kHz, and the time noise range to $\pm 0.02$. (a) Infidelity for standard optimization scheme. (b) Infidelity behavior where we use robust optimization with considering first-order ion-phonon coupling. (c) While considering both the first-order term of ion-phonon and ion-ion coupling, infidelity is robust for environmental noise.}
  \label{fig:robustrange}
\end{figure*}

\subsection{C. Robust for other noise}
To illustrate the effectiveness of our scheme for other kinds of noise, we also calculated the robustness of our scheme to the time noise or equivalent amplitude noise, where the time noise means the gate duration time change as $\tau \to \widetilde\tau$. As shown in Figure~\ref{fig:timenoise}, (a) illustrates that when we consider the higher order term of ion-phonon coupling, MS gate infidelity is sensitive to time noise, just as the normal optimization. While in our methods, after considering the higher-order term of ion-ion coupling, the optimized pulses are resistant to time noise. This result shows that in the quantum control of trapped ions, the higher-order term of the efficient coupling strength between ions plays a vital role in the realization of the multi-qubit gates. This effect can be seen intuitively in Figure~\ref{fig:timenoise} (b, c), in Figure~\ref{fig:timenoise} (b), we see that the residual entanglement between ion-phonons is robust to time noise after considering the higher-order terms of ion-phonons. However, in Figure~\ref{fig:timenoise} (c), when merely considering the higher ion-phonon coupling order term, we can observe that the efficient ion-ion coupling strength will soon depart from the maximum entanglement value as time noise increases.\\

Although in our cost function design, what we consider is the first derivative term of ion-phonon and ion-ion coupling to frequency drift, from Figure~\ref{fig:timenoise}, we have seen the effect of our scheme being robust to other noises as well. In Figure~\ref{fig:robustrange}, we numerically calculate the robustness properties of the MS gate under the combined influence of drift noise and time noise. As shown in Figure~\ref{fig:robustrange} (a), the optimized pulse is very sensitive to ambient noise by using a normal cost function. Whether it is drift noise or time noise, a little offset will cause a sharp drop in gate fidelity. After accounting for the higher-order term of ion-phonon coupling, this sensitivity improved, as shown in Figure~\ref{fig:robustrange} (b), but with a small range of robustness. If taking the higher-order term of ion-ion coupling into account, the optimized pulse has a drift noise range around $\pm 10$kHz and a timing noise around $\pm 0.02$, which can ensure that the error of the quantum gate below $10^{-3}$ as shown in \ref{fig:robustrange} (c). The improvement of the robustness range significantly promotes the realization of quantum error-correcting codes.

From Figure~\ref{fig:robustrange}, we can also notice a peculiar effect in the two dark regimes as shown in the Figure~\ref{fig:robustrange} (b, c), the optimized pulse can keep below $10^{-6}$ in a specific band. Along this noise response interval, perhaps we can set up positive feedback control by setting time offsets to reduce the effects of frequency drift noise, ensuring the accuracy of quantum operations. 

In summary, compared to the standard methods, our proposal dramatically improves the range of noise robustness by considering higher-order optimization. Furthermore, we note that our robust control is efficient for both frequency drift and other noise.

\section{V. Conclusion}
We have introduced a novel pulse design optimization method for trapped ions quantum control, and this method is motivated because gate fidelity endures many environmental noises in trapped ion platforms. By involving the higher-order term of ion-phonon and ion-ion coupling in the optimization function, we found that the ion-ion coupling is sensitive to the noise of the optical field. The higher-order coupling terms play a vital role in trapped ion quantum control optimization. We have analytically deduced a more suitable cost function for robust control optimization. To show the robustness of our methods, we have numerically performed the MS gate fidelity for a 4-fully connected trapped ions platform. The results show that our methods promote an admirable noise tolerance range, up to $\pm 10$kHz for drift noise and $\pm 0.02$ for time noise. Our work provides a beneficial pulse design scheme for realizing more efficient entangled states in trapped ion platforms, and we expect our method will improve the noise suppression in the experiment construction of trapped ion quantum computing.\\

\textit{\textbf{Note added.}} During completing this work, we became aware of two preprints appearing on arXiv recently~\cite{ruzic2022frequency, jia2022angle}, which also noticed the influence of higher-order ion-ion coupling expansion. These two works have realized a convincing experiment of a robust MS gate control. While work~\cite{ruzic2022frequency} experimental realizes the robustness of frequency drift noise by using Gaussian amplitude modulation, work~\cite{jia2022angle} experimental realizes robust MS gate using frequency modulation under frequency drift noise. Our work chooses an amplitude and phase hybrid modulation and numerically achieves robustness to frequency drift noise and time noise. The main results have been published on China Patent Application No.CN114707358A (in Chinese) since July 05, 2022~\cite{patent2022wang}.\\

\textit{\textbf{Acknowledgements.}} We want to thank Runyao Duan and Xin Wang for their helpful suggestions.

\bibliographystyle{apsrev4-1.bst}
\bibliography{trapped_ion_reference.bib}

\begin{thebibliography}{49}%
\makeatletter
\providecommand \@ifxundefined [1]{%
 \@ifx{#1\undefined}
}%
\providecommand \@ifnum [1]{%
 \ifnum #1\expandafter \@firstoftwo
 \else \expandafter \@secondoftwo
 \fi
}%
\providecommand \@ifx [1]{%
 \ifx #1\expandafter \@firstoftwo
 \else \expandafter \@secondoftwo
 \fi
}%
\providecommand \natexlab [1]{#1}%
\providecommand \enquote  [1]{``#1''}%
\providecommand \bibnamefont  [1]{#1}%
\providecommand \bibfnamefont [1]{#1}%
\providecommand \citenamefont [1]{#1}%
\providecommand \href@noop [0]{\@secondoftwo}%
\providecommand \href [0]{\begingroup \@sanitize@url \@href}%
\providecommand \@href[1]{\@@startlink{#1}\@@href}%
\providecommand \@@href[1]{\endgroup#1\@@endlink}%
\providecommand \@sanitize@url [0]{\catcode `\\12\catcode `\$12\catcode
  `\&12\catcode `\#12\catcode `\^12\catcode `\_12\catcode `\%12\relax}%
\providecommand \@@startlink[1]{}%
\providecommand \@@endlink[0]{}%
\providecommand \url  [0]{\begingroup\@sanitize@url \@url }%
\providecommand \@url [1]{\endgroup\@href {#1}{\urlprefix }}%
\providecommand \urlprefix  [0]{URL }%
\providecommand \Eprint [0]{\href }%
\providecommand \doibase [0]{http://dx.doi.org/}%
\providecommand \selectlanguage [0]{\@gobble}%
\providecommand \bibinfo  [0]{\@secondoftwo}%
\providecommand \bibfield  [0]{\@secondoftwo}%
\providecommand \translation [1]{[#1]}%
\providecommand \BibitemOpen [0]{}%
\providecommand \bibitemStop [0]{}%
\providecommand \bibitemNoStop [0]{.\EOS\space}%
\providecommand \EOS [0]{\spacefactor3000\relax}%
\providecommand \BibitemShut  [1]{\csname bibitem#1\endcsname}%
\let\auto@bib@innerbib\@empty
\bibitem [{\citenamefont {Bruzewicz}\ \emph {et~al.}(2019)\citenamefont
  {Bruzewicz}, \citenamefont {Chiaverini}, \citenamefont {McConnell},\ and\
  \citenamefont {Sage}}]{bruzewicz2019trapped}%
  \BibitemOpen
  \bibfield  {author} {\bibinfo {author} {\bibfnamefont {C.~D.}\ \bibnamefont
  {Bruzewicz}}, \bibinfo {author} {\bibfnamefont {J.}~\bibnamefont
  {Chiaverini}}, \bibinfo {author} {\bibfnamefont {R.}~\bibnamefont
  {McConnell}}, \ and\ \bibinfo {author} {\bibfnamefont {J.~M.}\ \bibnamefont
  {Sage}},\ }\href@noop {} {\bibfield  {journal} {\bibinfo  {journal} {Applied
  Physics Reviews}\ }\textbf {\bibinfo {volume} {6}},\ \bibinfo {pages}
  {021314} (\bibinfo {year} {2019})}\BibitemShut {NoStop}%
\bibitem [{\citenamefont {Mehta}\ \emph {et~al.}(2020)\citenamefont {Mehta},
  \citenamefont {Zhang}, \citenamefont {Malinowski}, \citenamefont {Nguyen},
  \citenamefont {Stadler},\ and\ \citenamefont {Home}}]{mehta2020integrated}%
  \BibitemOpen
  \bibfield  {author} {\bibinfo {author} {\bibfnamefont {K.~K.}\ \bibnamefont
  {Mehta}}, \bibinfo {author} {\bibfnamefont {C.}~\bibnamefont {Zhang}},
  \bibinfo {author} {\bibfnamefont {M.}~\bibnamefont {Malinowski}}, \bibinfo
  {author} {\bibfnamefont {T.-L.}\ \bibnamefont {Nguyen}}, \bibinfo {author}
  {\bibfnamefont {M.}~\bibnamefont {Stadler}}, \ and\ \bibinfo {author}
  {\bibfnamefont {J.~P.}\ \bibnamefont {Home}},\ }\href@noop {} {\bibfield
  {journal} {\bibinfo  {journal} {Nature}\ }\textbf {\bibinfo {volume} {586}},\
  \bibinfo {pages} {533} (\bibinfo {year} {2020})}\BibitemShut {NoStop}%
\bibitem [{\citenamefont {Niffenegger}\ \emph {et~al.}(2020)\citenamefont
  {Niffenegger}, \citenamefont {Stuart}, \citenamefont {Sorace-Agaskar},
  \citenamefont {Kharas}, \citenamefont {Bramhavar}, \citenamefont {Bruzewicz},
  \citenamefont {Loh}, \citenamefont {Maxson}, \citenamefont {McConnell},
  \citenamefont {Reens} \emph {et~al.}}]{niffenegger2020integrated}%
  \BibitemOpen
  \bibfield  {author} {\bibinfo {author} {\bibfnamefont {R.~J.}\ \bibnamefont
  {Niffenegger}}, \bibinfo {author} {\bibfnamefont {J.}~\bibnamefont {Stuart}},
  \bibinfo {author} {\bibfnamefont {C.}~\bibnamefont {Sorace-Agaskar}},
  \bibinfo {author} {\bibfnamefont {D.}~\bibnamefont {Kharas}}, \bibinfo
  {author} {\bibfnamefont {S.}~\bibnamefont {Bramhavar}}, \bibinfo {author}
  {\bibfnamefont {C.~D.}\ \bibnamefont {Bruzewicz}}, \bibinfo {author}
  {\bibfnamefont {W.}~\bibnamefont {Loh}}, \bibinfo {author} {\bibfnamefont
  {R.~T.}\ \bibnamefont {Maxson}}, \bibinfo {author} {\bibfnamefont
  {R.}~\bibnamefont {McConnell}}, \bibinfo {author} {\bibfnamefont
  {D.}~\bibnamefont {Reens}},  \emph {et~al.},\ }\href@noop {} {\bibfield
  {journal} {\bibinfo  {journal} {Nature}\ }\textbf {\bibinfo {volume} {586}},\
  \bibinfo {pages} {538} (\bibinfo {year} {2020})}\BibitemShut {NoStop}%
\bibitem [{\citenamefont {Grzesiak}\ \emph {et~al.}(2020)\citenamefont
  {Grzesiak}, \citenamefont {Bl{\"u}mel}, \citenamefont {Wright}, \citenamefont
  {Beck}, \citenamefont {Pisenti}, \citenamefont {Li}, \citenamefont {Chaplin},
  \citenamefont {Amini}, \citenamefont {Debnath}, \citenamefont {Chen} \emph
  {et~al.}}]{grzesiak2020efficient}%
  \BibitemOpen
  \bibfield  {author} {\bibinfo {author} {\bibfnamefont {N.}~\bibnamefont
  {Grzesiak}}, \bibinfo {author} {\bibfnamefont {R.}~\bibnamefont
  {Bl{\"u}mel}}, \bibinfo {author} {\bibfnamefont {K.}~\bibnamefont {Wright}},
  \bibinfo {author} {\bibfnamefont {K.~M.}\ \bibnamefont {Beck}}, \bibinfo
  {author} {\bibfnamefont {N.~C.}\ \bibnamefont {Pisenti}}, \bibinfo {author}
  {\bibfnamefont {M.}~\bibnamefont {Li}}, \bibinfo {author} {\bibfnamefont
  {V.}~\bibnamefont {Chaplin}}, \bibinfo {author} {\bibfnamefont {J.~M.}\
  \bibnamefont {Amini}}, \bibinfo {author} {\bibfnamefont {S.}~\bibnamefont
  {Debnath}}, \bibinfo {author} {\bibfnamefont {J.-S.}\ \bibnamefont {Chen}},
  \emph {et~al.},\ }\href@noop {} {\bibfield  {journal} {\bibinfo  {journal}
  {Nature communications}\ }\textbf {\bibinfo {volume} {11}},\ \bibinfo {pages}
  {1} (\bibinfo {year} {2020})}\BibitemShut {NoStop}%
\bibitem [{\citenamefont {Linke}\ \emph
  {et~al.}(2017{\natexlab{a}})\citenamefont {Linke}, \citenamefont {Maslov},
  \citenamefont {Roetteler}, \citenamefont {Debnath}, \citenamefont {Figgatt},
  \citenamefont {Landsman}, \citenamefont {Wright},\ and\ \citenamefont
  {Monroe}}]{linke2017experimental}%
  \BibitemOpen
  \bibfield  {author} {\bibinfo {author} {\bibfnamefont {N.~M.}\ \bibnamefont
  {Linke}}, \bibinfo {author} {\bibfnamefont {D.}~\bibnamefont {Maslov}},
  \bibinfo {author} {\bibfnamefont {M.}~\bibnamefont {Roetteler}}, \bibinfo
  {author} {\bibfnamefont {S.}~\bibnamefont {Debnath}}, \bibinfo {author}
  {\bibfnamefont {C.}~\bibnamefont {Figgatt}}, \bibinfo {author} {\bibfnamefont
  {K.~A.}\ \bibnamefont {Landsman}}, \bibinfo {author} {\bibfnamefont
  {K.}~\bibnamefont {Wright}}, \ and\ \bibinfo {author} {\bibfnamefont
  {C.}~\bibnamefont {Monroe}},\ }\href@noop {} {\bibfield  {journal} {\bibinfo
  {journal} {Proceedings of the National Academy of Sciences}\ }\textbf
  {\bibinfo {volume} {114}},\ \bibinfo {pages} {3305} (\bibinfo {year}
  {2017}{\natexlab{a}})}\BibitemShut {NoStop}%
\bibitem [{\citenamefont {Srinivas}\ \emph {et~al.}(2021)\citenamefont
  {Srinivas}, \citenamefont {Burd}, \citenamefont {Knaack}, \citenamefont
  {Sutherland}, \citenamefont {Kwiatkowski}, \citenamefont {Glancy},
  \citenamefont {Knill}, \citenamefont {Wineland}, \citenamefont {Leibfried},
  \citenamefont {Wilson} \emph {et~al.}}]{srinivas2021high}%
  \BibitemOpen
  \bibfield  {author} {\bibinfo {author} {\bibfnamefont {R.}~\bibnamefont
  {Srinivas}}, \bibinfo {author} {\bibfnamefont {S.}~\bibnamefont {Burd}},
  \bibinfo {author} {\bibfnamefont {H.}~\bibnamefont {Knaack}}, \bibinfo
  {author} {\bibfnamefont {R.}~\bibnamefont {Sutherland}}, \bibinfo {author}
  {\bibfnamefont {A.}~\bibnamefont {Kwiatkowski}}, \bibinfo {author}
  {\bibfnamefont {S.}~\bibnamefont {Glancy}}, \bibinfo {author} {\bibfnamefont
  {E.}~\bibnamefont {Knill}}, \bibinfo {author} {\bibfnamefont
  {D.}~\bibnamefont {Wineland}}, \bibinfo {author} {\bibfnamefont
  {D.}~\bibnamefont {Leibfried}}, \bibinfo {author} {\bibfnamefont {A.~C.}\
  \bibnamefont {Wilson}},  \emph {et~al.},\ }\href@noop {} {\bibfield
  {journal} {\bibinfo  {journal} {Nature}\ }\textbf {\bibinfo {volume} {597}},\
  \bibinfo {pages} {209} (\bibinfo {year} {2021})}\BibitemShut {NoStop}%
\bibitem [{\citenamefont {Ballance}\ \emph {et~al.}(2016)\citenamefont
  {Ballance}, \citenamefont {Harty}, \citenamefont {Linke}, \citenamefont
  {Sepiol},\ and\ \citenamefont {Lucas}}]{ballance2016high}%
  \BibitemOpen
  \bibfield  {author} {\bibinfo {author} {\bibfnamefont {C.}~\bibnamefont
  {Ballance}}, \bibinfo {author} {\bibfnamefont {T.}~\bibnamefont {Harty}},
  \bibinfo {author} {\bibfnamefont {N.}~\bibnamefont {Linke}}, \bibinfo
  {author} {\bibfnamefont {M.}~\bibnamefont {Sepiol}}, \ and\ \bibinfo {author}
  {\bibfnamefont {D.}~\bibnamefont {Lucas}},\ }\href@noop {} {\bibfield
  {journal} {\bibinfo  {journal} {Physical Review Letters}\ }\textbf {\bibinfo
  {volume} {117}},\ \bibinfo {pages} {060504} (\bibinfo {year}
  {2016})}\BibitemShut {NoStop}%
\bibitem [{\citenamefont {Gaebler}\ \emph {et~al.}(2016)\citenamefont
  {Gaebler}, \citenamefont {Tan}, \citenamefont {Lin}, \citenamefont {Wan},
  \citenamefont {Bowler}, \citenamefont {Keith}, \citenamefont {Glancy},
  \citenamefont {Coakley}, \citenamefont {Knill}, \citenamefont {Leibfried}
  \emph {et~al.}}]{gaebler2016high}%
  \BibitemOpen
  \bibfield  {author} {\bibinfo {author} {\bibfnamefont {J.~P.}\ \bibnamefont
  {Gaebler}}, \bibinfo {author} {\bibfnamefont {T.~R.}\ \bibnamefont {Tan}},
  \bibinfo {author} {\bibfnamefont {Y.}~\bibnamefont {Lin}}, \bibinfo {author}
  {\bibfnamefont {Y.}~\bibnamefont {Wan}}, \bibinfo {author} {\bibfnamefont
  {R.}~\bibnamefont {Bowler}}, \bibinfo {author} {\bibfnamefont {A.~C.}\
  \bibnamefont {Keith}}, \bibinfo {author} {\bibfnamefont {S.}~\bibnamefont
  {Glancy}}, \bibinfo {author} {\bibfnamefont {K.}~\bibnamefont {Coakley}},
  \bibinfo {author} {\bibfnamefont {E.}~\bibnamefont {Knill}}, \bibinfo
  {author} {\bibfnamefont {D.}~\bibnamefont {Leibfried}},  \emph {et~al.},\
  }\href@noop {} {\bibfield  {journal} {\bibinfo  {journal} {Physical Review
  Letters}\ }\textbf {\bibinfo {volume} {117}},\ \bibinfo {pages} {060505}
  (\bibinfo {year} {2016})}\BibitemShut {NoStop}%
\bibitem [{\citenamefont {Preskill}(2018)}]{preskill2018quantum}%
  \BibitemOpen
  \bibfield  {author} {\bibinfo {author} {\bibfnamefont {J.}~\bibnamefont
  {Preskill}},\ }\href@noop {} {\bibfield  {journal} {\bibinfo  {journal}
  {Quantum}\ }\textbf {\bibinfo {volume} {2}},\ \bibinfo {pages} {79} (\bibinfo
  {year} {2018})}\BibitemShut {NoStop}%
\bibitem [{\citenamefont {Zhang}\ \emph {et~al.}(2017)\citenamefont {Zhang},
  \citenamefont {Pagano}, \citenamefont {Hess}, \citenamefont {Kyprianidis},
  \citenamefont {Becker}, \citenamefont {Kaplan}, \citenamefont {Gorshkov},
  \citenamefont {Gong},\ and\ \citenamefont {Monroe}}]{zhang2017observation}%
  \BibitemOpen
  \bibfield  {author} {\bibinfo {author} {\bibfnamefont {J.}~\bibnamefont
  {Zhang}}, \bibinfo {author} {\bibfnamefont {G.}~\bibnamefont {Pagano}},
  \bibinfo {author} {\bibfnamefont {P.~W.}\ \bibnamefont {Hess}}, \bibinfo
  {author} {\bibfnamefont {A.}~\bibnamefont {Kyprianidis}}, \bibinfo {author}
  {\bibfnamefont {P.}~\bibnamefont {Becker}}, \bibinfo {author} {\bibfnamefont
  {H.}~\bibnamefont {Kaplan}}, \bibinfo {author} {\bibfnamefont {A.~V.}\
  \bibnamefont {Gorshkov}}, \bibinfo {author} {\bibfnamefont {Z.-X.}\
  \bibnamefont {Gong}}, \ and\ \bibinfo {author} {\bibfnamefont
  {C.}~\bibnamefont {Monroe}},\ }\href@noop {} {\bibfield  {journal} {\bibinfo
  {journal} {Nature}\ }\textbf {\bibinfo {volume} {551}},\ \bibinfo {pages}
  {601} (\bibinfo {year} {2017})}\BibitemShut {NoStop}%
\bibitem [{\citenamefont {Kawashima}\ \emph {et~al.}(2021)\citenamefont
  {Kawashima}, \citenamefont {Lloyd}, \citenamefont {Coons}, \citenamefont
  {Nam}, \citenamefont {Matsuura}, \citenamefont {Garza}, \citenamefont
  {Johri}, \citenamefont {Huntington}, \citenamefont {Senicourt}, \citenamefont
  {Maksymov} \emph {et~al.}}]{kawashima2021optimizing}%
  \BibitemOpen
  \bibfield  {author} {\bibinfo {author} {\bibfnamefont {Y.}~\bibnamefont
  {Kawashima}}, \bibinfo {author} {\bibfnamefont {E.}~\bibnamefont {Lloyd}},
  \bibinfo {author} {\bibfnamefont {M.~P.}\ \bibnamefont {Coons}}, \bibinfo
  {author} {\bibfnamefont {Y.}~\bibnamefont {Nam}}, \bibinfo {author}
  {\bibfnamefont {S.}~\bibnamefont {Matsuura}}, \bibinfo {author}
  {\bibfnamefont {A.~J.}\ \bibnamefont {Garza}}, \bibinfo {author}
  {\bibfnamefont {S.}~\bibnamefont {Johri}}, \bibinfo {author} {\bibfnamefont
  {L.}~\bibnamefont {Huntington}}, \bibinfo {author} {\bibfnamefont
  {V.}~\bibnamefont {Senicourt}}, \bibinfo {author} {\bibfnamefont {A.~O.}\
  \bibnamefont {Maksymov}},  \emph {et~al.},\ }\href@noop {} {\bibfield
  {journal} {\bibinfo  {journal} {Communications Physics}\ }\textbf {\bibinfo
  {volume} {4}},\ \bibinfo {pages} {1} (\bibinfo {year} {2021})}\BibitemShut
  {NoStop}%
\bibitem [{\citenamefont {Hempel}\ \emph {et~al.}(2018)\citenamefont {Hempel},
  \citenamefont {Maier}, \citenamefont {Romero}, \citenamefont {McClean},
  \citenamefont {Monz}, \citenamefont {Shen}, \citenamefont {Jurcevic},
  \citenamefont {Lanyon}, \citenamefont {Love}, \citenamefont {Babbush} \emph
  {et~al.}}]{hempel2018quantum}%
  \BibitemOpen
  \bibfield  {author} {\bibinfo {author} {\bibfnamefont {C.}~\bibnamefont
  {Hempel}}, \bibinfo {author} {\bibfnamefont {C.}~\bibnamefont {Maier}},
  \bibinfo {author} {\bibfnamefont {J.}~\bibnamefont {Romero}}, \bibinfo
  {author} {\bibfnamefont {J.}~\bibnamefont {McClean}}, \bibinfo {author}
  {\bibfnamefont {T.}~\bibnamefont {Monz}}, \bibinfo {author} {\bibfnamefont
  {H.}~\bibnamefont {Shen}}, \bibinfo {author} {\bibfnamefont {P.}~\bibnamefont
  {Jurcevic}}, \bibinfo {author} {\bibfnamefont {B.~P.}\ \bibnamefont
  {Lanyon}}, \bibinfo {author} {\bibfnamefont {P.}~\bibnamefont {Love}},
  \bibinfo {author} {\bibfnamefont {R.}~\bibnamefont {Babbush}},  \emph
  {et~al.},\ }\href@noop {} {\bibfield  {journal} {\bibinfo  {journal}
  {Physical Review X}\ }\textbf {\bibinfo {volume} {8}},\ \bibinfo {pages}
  {031022} (\bibinfo {year} {2018})}\BibitemShut {NoStop}%
\bibitem [{\citenamefont {Nam}\ \emph {et~al.}(2020)\citenamefont {Nam},
  \citenamefont {Chen}, \citenamefont {Pisenti}, \citenamefont {Wright},
  \citenamefont {Delaney}, \citenamefont {Maslov}, \citenamefont {Brown},
  \citenamefont {Allen}, \citenamefont {Amini}, \citenamefont {Apisdorf} \emph
  {et~al.}}]{nam2020ground}%
  \BibitemOpen
  \bibfield  {author} {\bibinfo {author} {\bibfnamefont {Y.}~\bibnamefont
  {Nam}}, \bibinfo {author} {\bibfnamefont {J.-S.}\ \bibnamefont {Chen}},
  \bibinfo {author} {\bibfnamefont {N.~C.}\ \bibnamefont {Pisenti}}, \bibinfo
  {author} {\bibfnamefont {K.}~\bibnamefont {Wright}}, \bibinfo {author}
  {\bibfnamefont {C.}~\bibnamefont {Delaney}}, \bibinfo {author} {\bibfnamefont
  {D.}~\bibnamefont {Maslov}}, \bibinfo {author} {\bibfnamefont {K.~R.}\
  \bibnamefont {Brown}}, \bibinfo {author} {\bibfnamefont {S.}~\bibnamefont
  {Allen}}, \bibinfo {author} {\bibfnamefont {J.~M.}\ \bibnamefont {Amini}},
  \bibinfo {author} {\bibfnamefont {J.}~\bibnamefont {Apisdorf}},  \emph
  {et~al.},\ }\href@noop {} {\bibfield  {journal} {\bibinfo  {journal} {npj
  Quantum Information}\ }\textbf {\bibinfo {volume} {6}},\ \bibinfo {pages} {1}
  (\bibinfo {year} {2020})}\BibitemShut {NoStop}%
\bibitem [{\citenamefont {Campbell}\ \emph {et~al.}(2017)\citenamefont
  {Campbell}, \citenamefont {Terhal},\ and\ \citenamefont
  {Vuillot}}]{campbell2017roads}%
  \BibitemOpen
  \bibfield  {author} {\bibinfo {author} {\bibfnamefont {E.~T.}\ \bibnamefont
  {Campbell}}, \bibinfo {author} {\bibfnamefont {B.~M.}\ \bibnamefont
  {Terhal}}, \ and\ \bibinfo {author} {\bibfnamefont {C.}~\bibnamefont
  {Vuillot}},\ }\href@noop {} {\bibfield  {journal} {\bibinfo  {journal}
  {Nature}\ }\textbf {\bibinfo {volume} {549}},\ \bibinfo {pages} {172}
  (\bibinfo {year} {2017})}\BibitemShut {NoStop}%
\bibitem [{\citenamefont {Linke}\ \emph
  {et~al.}(2017{\natexlab{b}})\citenamefont {Linke}, \citenamefont {Gutierrez},
  \citenamefont {Landsman}, \citenamefont {Figgatt}, \citenamefont {Debnath},
  \citenamefont {Brown},\ and\ \citenamefont {Monroe}}]{linke2017fault}%
  \BibitemOpen
  \bibfield  {author} {\bibinfo {author} {\bibfnamefont {N.~M.}\ \bibnamefont
  {Linke}}, \bibinfo {author} {\bibfnamefont {M.}~\bibnamefont {Gutierrez}},
  \bibinfo {author} {\bibfnamefont {K.~A.}\ \bibnamefont {Landsman}}, \bibinfo
  {author} {\bibfnamefont {C.}~\bibnamefont {Figgatt}}, \bibinfo {author}
  {\bibfnamefont {S.}~\bibnamefont {Debnath}}, \bibinfo {author} {\bibfnamefont
  {K.~R.}\ \bibnamefont {Brown}}, \ and\ \bibinfo {author} {\bibfnamefont
  {C.}~\bibnamefont {Monroe}},\ }\href@noop {} {\bibfield  {journal} {\bibinfo
  {journal} {Science advances}\ }\textbf {\bibinfo {volume} {3}},\ \bibinfo
  {pages} {e1701074} (\bibinfo {year} {2017}{\natexlab{b}})}\BibitemShut
  {NoStop}%
\bibitem [{\citenamefont {Bermudez}\ \emph {et~al.}(2017)\citenamefont
  {Bermudez}, \citenamefont {Xu}, \citenamefont {Nigmatullin}, \citenamefont
  {O’Gorman}, \citenamefont {Negnevitsky}, \citenamefont {Schindler},
  \citenamefont {Monz}, \citenamefont {Poschinger}, \citenamefont {Hempel},
  \citenamefont {Home} \emph {et~al.}}]{bermudez2017assessing}%
  \BibitemOpen
  \bibfield  {author} {\bibinfo {author} {\bibfnamefont {A.}~\bibnamefont
  {Bermudez}}, \bibinfo {author} {\bibfnamefont {X.}~\bibnamefont {Xu}},
  \bibinfo {author} {\bibfnamefont {R.}~\bibnamefont {Nigmatullin}}, \bibinfo
  {author} {\bibfnamefont {J.}~\bibnamefont {O’Gorman}}, \bibinfo {author}
  {\bibfnamefont {V.}~\bibnamefont {Negnevitsky}}, \bibinfo {author}
  {\bibfnamefont {P.}~\bibnamefont {Schindler}}, \bibinfo {author}
  {\bibfnamefont {T.}~\bibnamefont {Monz}}, \bibinfo {author} {\bibfnamefont
  {U.}~\bibnamefont {Poschinger}}, \bibinfo {author} {\bibfnamefont
  {C.}~\bibnamefont {Hempel}}, \bibinfo {author} {\bibfnamefont
  {J.}~\bibnamefont {Home}},  \emph {et~al.},\ }\href@noop {} {\bibfield
  {journal} {\bibinfo  {journal} {Physical Review X}\ }\textbf {\bibinfo
  {volume} {7}},\ \bibinfo {pages} {041061} (\bibinfo {year}
  {2017})}\BibitemShut {NoStop}%
\bibitem [{\citenamefont {Benhelm}\ \emph {et~al.}(2008)\citenamefont
  {Benhelm}, \citenamefont {Kirchmair}, \citenamefont {Roos},\ and\
  \citenamefont {Blatt}}]{benhelm2008towards}%
  \BibitemOpen
  \bibfield  {author} {\bibinfo {author} {\bibfnamefont {J.}~\bibnamefont
  {Benhelm}}, \bibinfo {author} {\bibfnamefont {G.}~\bibnamefont {Kirchmair}},
  \bibinfo {author} {\bibfnamefont {C.~F.}\ \bibnamefont {Roos}}, \ and\
  \bibinfo {author} {\bibfnamefont {R.}~\bibnamefont {Blatt}},\ }\href@noop {}
  {\bibfield  {journal} {\bibinfo  {journal} {Nature Physics}\ }\textbf
  {\bibinfo {volume} {4}},\ \bibinfo {pages} {463} (\bibinfo {year}
  {2008})}\BibitemShut {NoStop}%
\bibitem [{\citenamefont {Paul}(1990)}]{paul1990electromagnetic}%
  \BibitemOpen
  \bibfield  {author} {\bibinfo {author} {\bibfnamefont {W.}~\bibnamefont
  {Paul}},\ }\href@noop {} {\bibfield  {journal} {\bibinfo  {journal} {Reviews
  of modern physics}\ }\textbf {\bibinfo {volume} {62}},\ \bibinfo {pages}
  {531} (\bibinfo {year} {1990})}\BibitemShut {NoStop}%
\bibitem [{\citenamefont {Brown}\ and\ \citenamefont
  {Gabrielse}(1986)}]{brown1986geonium}%
  \BibitemOpen
  \bibfield  {author} {\bibinfo {author} {\bibfnamefont {L.~S.}\ \bibnamefont
  {Brown}}\ and\ \bibinfo {author} {\bibfnamefont {G.}~\bibnamefont
  {Gabrielse}},\ }\href@noop {} {\bibfield  {journal} {\bibinfo  {journal}
  {Reviews of Modern Physics}\ }\textbf {\bibinfo {volume} {58}},\ \bibinfo
  {pages} {233} (\bibinfo {year} {1986})}\BibitemShut {NoStop}%
\bibitem [{\citenamefont {Britton}\ \emph {et~al.}(2012)\citenamefont
  {Britton}, \citenamefont {Sawyer}, \citenamefont {Keith}, \citenamefont
  {Wang}, \citenamefont {Freericks}, \citenamefont {Uys}, \citenamefont
  {Biercuk},\ and\ \citenamefont {Bollinger}}]{britton2012engineered}%
  \BibitemOpen
  \bibfield  {author} {\bibinfo {author} {\bibfnamefont {J.~W.}\ \bibnamefont
  {Britton}}, \bibinfo {author} {\bibfnamefont {B.~C.}\ \bibnamefont {Sawyer}},
  \bibinfo {author} {\bibfnamefont {A.~C.}\ \bibnamefont {Keith}}, \bibinfo
  {author} {\bibfnamefont {C.-C.~J.}\ \bibnamefont {Wang}}, \bibinfo {author}
  {\bibfnamefont {J.~K.}\ \bibnamefont {Freericks}}, \bibinfo {author}
  {\bibfnamefont {H.}~\bibnamefont {Uys}}, \bibinfo {author} {\bibfnamefont
  {M.~J.}\ \bibnamefont {Biercuk}}, \ and\ \bibinfo {author} {\bibfnamefont
  {J.~J.}\ \bibnamefont {Bollinger}},\ }\href@noop {} {\bibfield  {journal}
  {\bibinfo  {journal} {Nature}\ }\textbf {\bibinfo {volume} {484}},\ \bibinfo
  {pages} {489} (\bibinfo {year} {2012})}\BibitemShut {NoStop}%
\bibitem [{\citenamefont {Kielpinski}\ \emph {et~al.}(2002)\citenamefont
  {Kielpinski}, \citenamefont {Monroe},\ and\ \citenamefont
  {Wineland}}]{kielpinski2002architecture}%
  \BibitemOpen
  \bibfield  {author} {\bibinfo {author} {\bibfnamefont {D.}~\bibnamefont
  {Kielpinski}}, \bibinfo {author} {\bibfnamefont {C.}~\bibnamefont {Monroe}},
  \ and\ \bibinfo {author} {\bibfnamefont {D.~J.}\ \bibnamefont {Wineland}},\
  }\href@noop {} {\bibfield  {journal} {\bibinfo  {journal} {Nature}\ }\textbf
  {\bibinfo {volume} {417}},\ \bibinfo {pages} {709} (\bibinfo {year}
  {2002})}\BibitemShut {NoStop}%
\bibitem [{\citenamefont {Bowler}\ \emph {et~al.}(2012)\citenamefont {Bowler},
  \citenamefont {Gaebler}, \citenamefont {Lin}, \citenamefont {Tan},
  \citenamefont {Hanneke}, \citenamefont {Jost}, \citenamefont {Home},
  \citenamefont {Leibfried},\ and\ \citenamefont
  {Wineland}}]{bowler2012coherent}%
  \BibitemOpen
  \bibfield  {author} {\bibinfo {author} {\bibfnamefont {R.}~\bibnamefont
  {Bowler}}, \bibinfo {author} {\bibfnamefont {J.}~\bibnamefont {Gaebler}},
  \bibinfo {author} {\bibfnamefont {Y.}~\bibnamefont {Lin}}, \bibinfo {author}
  {\bibfnamefont {T.~R.}\ \bibnamefont {Tan}}, \bibinfo {author} {\bibfnamefont
  {D.}~\bibnamefont {Hanneke}}, \bibinfo {author} {\bibfnamefont {J.~D.}\
  \bibnamefont {Jost}}, \bibinfo {author} {\bibfnamefont {J.}~\bibnamefont
  {Home}}, \bibinfo {author} {\bibfnamefont {D.}~\bibnamefont {Leibfried}}, \
  and\ \bibinfo {author} {\bibfnamefont {D.~J.}\ \bibnamefont {Wineland}},\
  }\href@noop {} {\bibfield  {journal} {\bibinfo  {journal} {Physical Review
  Letters}\ }\textbf {\bibinfo {volume} {109}},\ \bibinfo {pages} {080502}
  (\bibinfo {year} {2012})}\BibitemShut {NoStop}%
\bibitem [{\citenamefont {Pino}\ \emph {et~al.}(2021)\citenamefont {Pino},
  \citenamefont {Dreiling}, \citenamefont {Figgatt}, \citenamefont {Gaebler},
  \citenamefont {Moses}, \citenamefont {Allman}, \citenamefont {Baldwin},
  \citenamefont {Foss-Feig}, \citenamefont {Hayes}, \citenamefont {Mayer} \emph
  {et~al.}}]{pino2021demonstration}%
  \BibitemOpen
  \bibfield  {author} {\bibinfo {author} {\bibfnamefont {J.~M.}\ \bibnamefont
  {Pino}}, \bibinfo {author} {\bibfnamefont {J.~M.}\ \bibnamefont {Dreiling}},
  \bibinfo {author} {\bibfnamefont {C.}~\bibnamefont {Figgatt}}, \bibinfo
  {author} {\bibfnamefont {J.~P.}\ \bibnamefont {Gaebler}}, \bibinfo {author}
  {\bibfnamefont {S.~A.}\ \bibnamefont {Moses}}, \bibinfo {author}
  {\bibfnamefont {M.}~\bibnamefont {Allman}}, \bibinfo {author} {\bibfnamefont
  {C.}~\bibnamefont {Baldwin}}, \bibinfo {author} {\bibfnamefont
  {M.}~\bibnamefont {Foss-Feig}}, \bibinfo {author} {\bibfnamefont
  {D.}~\bibnamefont {Hayes}}, \bibinfo {author} {\bibfnamefont
  {K.}~\bibnamefont {Mayer}},  \emph {et~al.},\ }\href@noop {} {\bibfield
  {journal} {\bibinfo  {journal} {Nature}\ }\textbf {\bibinfo {volume} {592}},\
  \bibinfo {pages} {209} (\bibinfo {year} {2021})}\BibitemShut {NoStop}%
\bibitem [{\citenamefont {Choi}\ \emph {et~al.}(2014)\citenamefont {Choi},
  \citenamefont {Debnath}, \citenamefont {Manning}, \citenamefont {Figgatt},
  \citenamefont {Gong}, \citenamefont {Duan},\ and\ \citenamefont
  {Monroe}}]{choi2014optimal}%
  \BibitemOpen
  \bibfield  {author} {\bibinfo {author} {\bibfnamefont {T.}~\bibnamefont
  {Choi}}, \bibinfo {author} {\bibfnamefont {S.}~\bibnamefont {Debnath}},
  \bibinfo {author} {\bibfnamefont {T.}~\bibnamefont {Manning}}, \bibinfo
  {author} {\bibfnamefont {C.}~\bibnamefont {Figgatt}}, \bibinfo {author}
  {\bibfnamefont {Z.-X.}\ \bibnamefont {Gong}}, \bibinfo {author}
  {\bibfnamefont {L.-M.}\ \bibnamefont {Duan}}, \ and\ \bibinfo {author}
  {\bibfnamefont {C.}~\bibnamefont {Monroe}},\ }\href@noop {} {\bibfield
  {journal} {\bibinfo  {journal} {Physical Review Letters}\ }\textbf {\bibinfo
  {volume} {112}},\ \bibinfo {pages} {190502} (\bibinfo {year}
  {2014})}\BibitemShut {NoStop}%
\bibitem [{\citenamefont {Myerson}\ \emph {et~al.}(2008)\citenamefont
  {Myerson}, \citenamefont {Szwer}, \citenamefont {Webster}, \citenamefont
  {Allcock}, \citenamefont {Curtis}, \citenamefont {Imreh}, \citenamefont
  {Sherman}, \citenamefont {Stacey}, \citenamefont {Steane},\ and\
  \citenamefont {Lucas}}]{myerson2008high}%
  \BibitemOpen
  \bibfield  {author} {\bibinfo {author} {\bibfnamefont {A.}~\bibnamefont
  {Myerson}}, \bibinfo {author} {\bibfnamefont {D.}~\bibnamefont {Szwer}},
  \bibinfo {author} {\bibfnamefont {S.}~\bibnamefont {Webster}}, \bibinfo
  {author} {\bibfnamefont {D.}~\bibnamefont {Allcock}}, \bibinfo {author}
  {\bibfnamefont {M.}~\bibnamefont {Curtis}}, \bibinfo {author} {\bibfnamefont
  {G.}~\bibnamefont {Imreh}}, \bibinfo {author} {\bibfnamefont
  {J.}~\bibnamefont {Sherman}}, \bibinfo {author} {\bibfnamefont
  {D.}~\bibnamefont {Stacey}}, \bibinfo {author} {\bibfnamefont
  {A.}~\bibnamefont {Steane}}, \ and\ \bibinfo {author} {\bibfnamefont
  {D.}~\bibnamefont {Lucas}},\ }\href@noop {} {\bibfield  {journal} {\bibinfo
  {journal} {Physical Review Letters}\ }\textbf {\bibinfo {volume} {100}},\
  \bibinfo {pages} {200502} (\bibinfo {year} {2008})}\BibitemShut {NoStop}%
\bibitem [{\citenamefont {Harty}\ \emph {et~al.}(2014)\citenamefont {Harty},
  \citenamefont {Allcock}, \citenamefont {Ballance}, \citenamefont {Guidoni},
  \citenamefont {Janacek}, \citenamefont {Linke}, \citenamefont {Stacey},\ and\
  \citenamefont {Lucas}}]{harty2014high}%
  \BibitemOpen
  \bibfield  {author} {\bibinfo {author} {\bibfnamefont {T.}~\bibnamefont
  {Harty}}, \bibinfo {author} {\bibfnamefont {D.}~\bibnamefont {Allcock}},
  \bibinfo {author} {\bibfnamefont {C.~J.}\ \bibnamefont {Ballance}}, \bibinfo
  {author} {\bibfnamefont {L.}~\bibnamefont {Guidoni}}, \bibinfo {author}
  {\bibfnamefont {H.}~\bibnamefont {Janacek}}, \bibinfo {author} {\bibfnamefont
  {N.}~\bibnamefont {Linke}}, \bibinfo {author} {\bibfnamefont
  {D.}~\bibnamefont {Stacey}}, \ and\ \bibinfo {author} {\bibfnamefont
  {D.}~\bibnamefont {Lucas}},\ }\href@noop {} {\bibfield  {journal} {\bibinfo
  {journal} {Physical Review Letters}\ }\textbf {\bibinfo {volume} {113}},\
  \bibinfo {pages} {220501} (\bibinfo {year} {2014})}\BibitemShut {NoStop}%
\bibitem [{\citenamefont {Wu}\ \emph {et~al.}(2018)\citenamefont {Wu},
  \citenamefont {Wang},\ and\ \citenamefont {Duan}}]{wu2018noise}%
  \BibitemOpen
  \bibfield  {author} {\bibinfo {author} {\bibfnamefont {Y.}~\bibnamefont
  {Wu}}, \bibinfo {author} {\bibfnamefont {S.-T.}\ \bibnamefont {Wang}}, \ and\
  \bibinfo {author} {\bibfnamefont {L.-M.}\ \bibnamefont {Duan}},\ }\href@noop
  {} {\bibfield  {journal} {\bibinfo  {journal} {Physical Review A}\ }\textbf
  {\bibinfo {volume} {97}},\ \bibinfo {pages} {062325} (\bibinfo {year}
  {2018})}\BibitemShut {NoStop}%
\bibitem [{\citenamefont {Hayes}\ \emph {et~al.}(2012)\citenamefont {Hayes},
  \citenamefont {Clark}, \citenamefont {Debnath}, \citenamefont {Hucul},
  \citenamefont {Inlek}, \citenamefont {Lee}, \citenamefont {Quraishi},\ and\
  \citenamefont {Monroe}}]{hayes2012coherent}%
  \BibitemOpen
  \bibfield  {author} {\bibinfo {author} {\bibfnamefont {D.}~\bibnamefont
  {Hayes}}, \bibinfo {author} {\bibfnamefont {S.~M.}\ \bibnamefont {Clark}},
  \bibinfo {author} {\bibfnamefont {S.}~\bibnamefont {Debnath}}, \bibinfo
  {author} {\bibfnamefont {D.}~\bibnamefont {Hucul}}, \bibinfo {author}
  {\bibfnamefont {I.~V.}\ \bibnamefont {Inlek}}, \bibinfo {author}
  {\bibfnamefont {K.~W.}\ \bibnamefont {Lee}}, \bibinfo {author} {\bibfnamefont
  {Q.}~\bibnamefont {Quraishi}}, \ and\ \bibinfo {author} {\bibfnamefont
  {C.}~\bibnamefont {Monroe}},\ }\href@noop {} {\bibfield  {journal} {\bibinfo
  {journal} {Physical Review Letters}\ }\textbf {\bibinfo {volume} {109}},\
  \bibinfo {pages} {020503} (\bibinfo {year} {2012})}\BibitemShut {NoStop}%
\bibitem [{\citenamefont {Haddadfarshi}\ and\ \citenamefont
  {Mintert}(2016)}]{haddadfarshi2016high}%
  \BibitemOpen
  \bibfield  {author} {\bibinfo {author} {\bibfnamefont {F.}~\bibnamefont
  {Haddadfarshi}}\ and\ \bibinfo {author} {\bibfnamefont {F.}~\bibnamefont
  {Mintert}},\ }\href@noop {} {\bibfield  {journal} {\bibinfo  {journal} {New
  Journal of Physics}\ }\textbf {\bibinfo {volume} {18}},\ \bibinfo {pages}
  {123007} (\bibinfo {year} {2016})}\BibitemShut {NoStop}%
\bibitem [{\citenamefont {Manovitz}\ \emph {et~al.}(2017)\citenamefont
  {Manovitz}, \citenamefont {Rotem}, \citenamefont {Shaniv}, \citenamefont
  {Cohen}, \citenamefont {Shapira}, \citenamefont {Akerman}, \citenamefont
  {Retzker},\ and\ \citenamefont {Ozeri}}]{manovitz2017fast}%
  \BibitemOpen
  \bibfield  {author} {\bibinfo {author} {\bibfnamefont {T.}~\bibnamefont
  {Manovitz}}, \bibinfo {author} {\bibfnamefont {A.}~\bibnamefont {Rotem}},
  \bibinfo {author} {\bibfnamefont {R.}~\bibnamefont {Shaniv}}, \bibinfo
  {author} {\bibfnamefont {I.}~\bibnamefont {Cohen}}, \bibinfo {author}
  {\bibfnamefont {Y.}~\bibnamefont {Shapira}}, \bibinfo {author} {\bibfnamefont
  {N.}~\bibnamefont {Akerman}}, \bibinfo {author} {\bibfnamefont
  {A.}~\bibnamefont {Retzker}}, \ and\ \bibinfo {author} {\bibfnamefont
  {R.}~\bibnamefont {Ozeri}},\ }\href@noop {} {\bibfield  {journal} {\bibinfo
  {journal} {Physical Review Letters}\ }\textbf {\bibinfo {volume} {119}},\
  \bibinfo {pages} {220505} (\bibinfo {year} {2017})}\BibitemShut {NoStop}%
\bibitem [{\citenamefont {Shapira}\ \emph {et~al.}(2018)\citenamefont
  {Shapira}, \citenamefont {Shaniv}, \citenamefont {Manovitz}, \citenamefont
  {Akerman},\ and\ \citenamefont {Ozeri}}]{shapira2018robust}%
  \BibitemOpen
  \bibfield  {author} {\bibinfo {author} {\bibfnamefont {Y.}~\bibnamefont
  {Shapira}}, \bibinfo {author} {\bibfnamefont {R.}~\bibnamefont {Shaniv}},
  \bibinfo {author} {\bibfnamefont {T.}~\bibnamefont {Manovitz}}, \bibinfo
  {author} {\bibfnamefont {N.}~\bibnamefont {Akerman}}, \ and\ \bibinfo
  {author} {\bibfnamefont {R.}~\bibnamefont {Ozeri}},\ }\href@noop {}
  {\bibfield  {journal} {\bibinfo  {journal} {Physical Review Letters}\
  }\textbf {\bibinfo {volume} {121}},\ \bibinfo {pages} {180502} (\bibinfo
  {year} {2018})}\BibitemShut {NoStop}%
\bibitem [{\citenamefont {Webb}\ \emph {et~al.}(2018)\citenamefont {Webb},
  \citenamefont {Webster}, \citenamefont {Collingbourne}, \citenamefont
  {Bretaud}, \citenamefont {Lawrence}, \citenamefont {Weidt}, \citenamefont
  {Mintert},\ and\ \citenamefont {Hensinger}}]{webb2018resilient}%
  \BibitemOpen
  \bibfield  {author} {\bibinfo {author} {\bibfnamefont {A.~E.}\ \bibnamefont
  {Webb}}, \bibinfo {author} {\bibfnamefont {S.~C.}\ \bibnamefont {Webster}},
  \bibinfo {author} {\bibfnamefont {S.}~\bibnamefont {Collingbourne}}, \bibinfo
  {author} {\bibfnamefont {D.}~\bibnamefont {Bretaud}}, \bibinfo {author}
  {\bibfnamefont {A.~M.}\ \bibnamefont {Lawrence}}, \bibinfo {author}
  {\bibfnamefont {S.}~\bibnamefont {Weidt}}, \bibinfo {author} {\bibfnamefont
  {F.}~\bibnamefont {Mintert}}, \ and\ \bibinfo {author} {\bibfnamefont
  {W.~K.}\ \bibnamefont {Hensinger}},\ }\href@noop {} {\bibfield  {journal}
  {\bibinfo  {journal} {Physical Review Letters}\ }\textbf {\bibinfo {volume}
  {121}},\ \bibinfo {pages} {180501} (\bibinfo {year} {2018})}\BibitemShut
  {NoStop}%
\bibitem [{\citenamefont {Zarantonello}\ \emph {et~al.}(2019)\citenamefont
  {Zarantonello}, \citenamefont {Hahn}, \citenamefont {Morgner}, \citenamefont
  {Schulte}, \citenamefont {Bautista-Salvador}, \citenamefont {Werner},
  \citenamefont {Hammerer},\ and\ \citenamefont
  {Ospelkaus}}]{zarantonello2019robust}%
  \BibitemOpen
  \bibfield  {author} {\bibinfo {author} {\bibfnamefont {G.}~\bibnamefont
  {Zarantonello}}, \bibinfo {author} {\bibfnamefont {H.}~\bibnamefont {Hahn}},
  \bibinfo {author} {\bibfnamefont {J.}~\bibnamefont {Morgner}}, \bibinfo
  {author} {\bibfnamefont {M.}~\bibnamefont {Schulte}}, \bibinfo {author}
  {\bibfnamefont {A.}~\bibnamefont {Bautista-Salvador}}, \bibinfo {author}
  {\bibfnamefont {R.}~\bibnamefont {Werner}}, \bibinfo {author} {\bibfnamefont
  {K.}~\bibnamefont {Hammerer}}, \ and\ \bibinfo {author} {\bibfnamefont
  {C.}~\bibnamefont {Ospelkaus}},\ }\href@noop {} {\bibfield  {journal}
  {\bibinfo  {journal} {Physical Review Letters}\ }\textbf {\bibinfo {volume}
  {123}},\ \bibinfo {pages} {260503} (\bibinfo {year} {2019})}\BibitemShut
  {NoStop}%
\bibitem [{\citenamefont {Ruzic}\ \emph {et~al.}(2022)\citenamefont {Ruzic},
  \citenamefont {Chow}, \citenamefont {Burch}, \citenamefont {Lobser},
  \citenamefont {Revelle}, \citenamefont {Wilson}, \citenamefont {Yale},\ and\
  \citenamefont {Clark}}]{ruzic2022frequency}%
  \BibitemOpen
  \bibfield  {author} {\bibinfo {author} {\bibfnamefont {B.~P.}\ \bibnamefont
  {Ruzic}}, \bibinfo {author} {\bibfnamefont {M.~N.}\ \bibnamefont {Chow}},
  \bibinfo {author} {\bibfnamefont {A.~D.}\ \bibnamefont {Burch}}, \bibinfo
  {author} {\bibfnamefont {D.}~\bibnamefont {Lobser}}, \bibinfo {author}
  {\bibfnamefont {M.~C.}\ \bibnamefont {Revelle}}, \bibinfo {author}
  {\bibfnamefont {J.~M.}\ \bibnamefont {Wilson}}, \bibinfo {author}
  {\bibfnamefont {C.~G.}\ \bibnamefont {Yale}}, \ and\ \bibinfo {author}
  {\bibfnamefont {S.~M.}\ \bibnamefont {Clark}},\ }\href@noop {} {\bibfield
  {journal} {\bibinfo  {journal} {arXiv preprint arXiv:2210.02372}\ } (\bibinfo
  {year} {2022})}\BibitemShut {NoStop}%
\bibitem [{\citenamefont {Jia}\ \emph {et~al.}(2022)\citenamefont {Jia},
  \citenamefont {Huang}, \citenamefont {Kang}, \citenamefont {Sun},
  \citenamefont {Spivey}, \citenamefont {Kim},\ and\ \citenamefont
  {Brown}}]{jia2022angle}%
  \BibitemOpen
  \bibfield  {author} {\bibinfo {author} {\bibfnamefont {Z.}~\bibnamefont
  {Jia}}, \bibinfo {author} {\bibfnamefont {S.}~\bibnamefont {Huang}}, \bibinfo
  {author} {\bibfnamefont {M.}~\bibnamefont {Kang}}, \bibinfo {author}
  {\bibfnamefont {K.}~\bibnamefont {Sun}}, \bibinfo {author} {\bibfnamefont
  {R.~F.}\ \bibnamefont {Spivey}}, \bibinfo {author} {\bibfnamefont
  {J.}~\bibnamefont {Kim}}, \ and\ \bibinfo {author} {\bibfnamefont {K.~R.}\
  \bibnamefont {Brown}},\ }\href@noop {} {\bibfield  {journal} {\bibinfo
  {journal} {arXiv preprint arXiv:2210.04814}\ } (\bibinfo {year}
  {2022})}\BibitemShut {NoStop}%
\bibitem [{\citenamefont {Nielsen}\ and\ \citenamefont
  {Chuang}(2002)}]{nielsen2002quantum}%
  \BibitemOpen
  \bibfield  {author} {\bibinfo {author} {\bibfnamefont {M.~A.}\ \bibnamefont
  {Nielsen}}\ and\ \bibinfo {author} {\bibfnamefont {I.}~\bibnamefont
  {Chuang}},\ }\href@noop {} {\enquote {\bibinfo {title} {Quantum computation
  and quantum information},}\ } (\bibinfo {year} {2002})\BibitemShut {NoStop}%
\bibitem [{\citenamefont {S{\o}rensen}\ and\ \citenamefont
  {M{\o}lmer}(1999)}]{sorensen1999quantum}%
  \BibitemOpen
  \bibfield  {author} {\bibinfo {author} {\bibfnamefont {A.}~\bibnamefont
  {S{\o}rensen}}\ and\ \bibinfo {author} {\bibfnamefont {K.}~\bibnamefont
  {M{\o}lmer}},\ }\href@noop {} {\bibfield  {journal} {\bibinfo  {journal}
  {Physical Review Letters}\ }\textbf {\bibinfo {volume} {82}},\ \bibinfo
  {pages} {1971} (\bibinfo {year} {1999})}\BibitemShut {NoStop}%
\bibitem [{\citenamefont {Zhu}\ \emph {et~al.}(2006)\citenamefont {Zhu},
  \citenamefont {Monroe},\ and\ \citenamefont {Duan}}]{zhu2006arbitrary}%
  \BibitemOpen
  \bibfield  {author} {\bibinfo {author} {\bibfnamefont {S.-L.}\ \bibnamefont
  {Zhu}}, \bibinfo {author} {\bibfnamefont {C.}~\bibnamefont {Monroe}}, \ and\
  \bibinfo {author} {\bibfnamefont {L.-M.}\ \bibnamefont {Duan}},\ }\href@noop
  {} {\bibfield  {journal} {\bibinfo  {journal} {EPL (Europhysics Letters)}\
  }\textbf {\bibinfo {volume} {73}},\ \bibinfo {pages} {485} (\bibinfo {year}
  {2006})}\BibitemShut {NoStop}%
\bibitem [{\citenamefont {Roos}(2008)}]{roos2008ion}%
  \BibitemOpen
  \bibfield  {author} {\bibinfo {author} {\bibfnamefont {C.~F.}\ \bibnamefont
  {Roos}},\ }\href@noop {} {\bibfield  {journal} {\bibinfo  {journal} {New
  Journal of Physics}\ }\textbf {\bibinfo {volume} {10}},\ \bibinfo {pages}
  {013002} (\bibinfo {year} {2008})}\BibitemShut {NoStop}%
\bibitem [{\citenamefont {Steane}\ \emph {et~al.}(2014)\citenamefont {Steane},
  \citenamefont {Imreh}, \citenamefont {Home},\ and\ \citenamefont
  {Leibfried}}]{steane2014pulsed}%
  \BibitemOpen
  \bibfield  {author} {\bibinfo {author} {\bibfnamefont {A.~M.}\ \bibnamefont
  {Steane}}, \bibinfo {author} {\bibfnamefont {G.}~\bibnamefont {Imreh}},
  \bibinfo {author} {\bibfnamefont {J.~P.}\ \bibnamefont {Home}}, \ and\
  \bibinfo {author} {\bibfnamefont {D.}~\bibnamefont {Leibfried}},\ }\href@noop
  {} {\bibfield  {journal} {\bibinfo  {journal} {New Journal of Physics}\
  }\textbf {\bibinfo {volume} {16}},\ \bibinfo {pages} {053049} (\bibinfo
  {year} {2014})}\BibitemShut {NoStop}%
\bibitem [{\citenamefont {Leung}\ \emph {et~al.}(2018)\citenamefont {Leung},
  \citenamefont {Landsman}, \citenamefont {Figgatt}, \citenamefont {Linke},
  \citenamefont {Monroe},\ and\ \citenamefont {Brown}}]{leung2018robust}%
  \BibitemOpen
  \bibfield  {author} {\bibinfo {author} {\bibfnamefont {P.~H.}\ \bibnamefont
  {Leung}}, \bibinfo {author} {\bibfnamefont {K.~A.}\ \bibnamefont {Landsman}},
  \bibinfo {author} {\bibfnamefont {C.}~\bibnamefont {Figgatt}}, \bibinfo
  {author} {\bibfnamefont {N.~M.}\ \bibnamefont {Linke}}, \bibinfo {author}
  {\bibfnamefont {C.}~\bibnamefont {Monroe}}, \ and\ \bibinfo {author}
  {\bibfnamefont {K.~R.}\ \bibnamefont {Brown}},\ }\href@noop {} {\bibfield
  {journal} {\bibinfo  {journal} {Physical Review Letters}\ }\textbf {\bibinfo
  {volume} {120}},\ \bibinfo {pages} {020501} (\bibinfo {year}
  {2018})}\BibitemShut {NoStop}%
\bibitem [{\citenamefont {Leung}\ and\ \citenamefont
  {Brown}(2018)}]{leung2018entangling}%
  \BibitemOpen
  \bibfield  {author} {\bibinfo {author} {\bibfnamefont {P.~H.}\ \bibnamefont
  {Leung}}\ and\ \bibinfo {author} {\bibfnamefont {K.~R.}\ \bibnamefont
  {Brown}},\ }\href@noop {} {\bibfield  {journal} {\bibinfo  {journal}
  {Physical Review A}\ }\textbf {\bibinfo {volume} {98}},\ \bibinfo {pages}
  {032318} (\bibinfo {year} {2018})}\BibitemShut {NoStop}%
\bibitem [{\citenamefont {Lu}\ \emph {et~al.}(2019)\citenamefont {Lu},
  \citenamefont {Zhang}, \citenamefont {Zhang}, \citenamefont {Chen},
  \citenamefont {Shen}, \citenamefont {Zhang}, \citenamefont {Zhang},\ and\
  \citenamefont {Kim}}]{lu2019global}%
  \BibitemOpen
  \bibfield  {author} {\bibinfo {author} {\bibfnamefont {Y.}~\bibnamefont
  {Lu}}, \bibinfo {author} {\bibfnamefont {S.}~\bibnamefont {Zhang}}, \bibinfo
  {author} {\bibfnamefont {K.}~\bibnamefont {Zhang}}, \bibinfo {author}
  {\bibfnamefont {W.}~\bibnamefont {Chen}}, \bibinfo {author} {\bibfnamefont
  {Y.}~\bibnamefont {Shen}}, \bibinfo {author} {\bibfnamefont {J.}~\bibnamefont
  {Zhang}}, \bibinfo {author} {\bibfnamefont {J.-N.}\ \bibnamefont {Zhang}}, \
  and\ \bibinfo {author} {\bibfnamefont {K.}~\bibnamefont {Kim}},\ }\href@noop
  {} {\bibfield  {journal} {\bibinfo  {journal} {Nature}\ }\textbf {\bibinfo
  {volume} {572}},\ \bibinfo {pages} {363} (\bibinfo {year}
  {2019})}\BibitemShut {NoStop}%
\bibitem [{\citenamefont {Landsman}\ \emph {et~al.}(2019)\citenamefont
  {Landsman}, \citenamefont {Wu}, \citenamefont {Leung}, \citenamefont {Zhu},
  \citenamefont {Linke}, \citenamefont {Brown}, \citenamefont {Duan},\ and\
  \citenamefont {Monroe}}]{landsman2019two}%
  \BibitemOpen
  \bibfield  {author} {\bibinfo {author} {\bibfnamefont {K.~A.}\ \bibnamefont
  {Landsman}}, \bibinfo {author} {\bibfnamefont {Y.}~\bibnamefont {Wu}},
  \bibinfo {author} {\bibfnamefont {P.~H.}\ \bibnamefont {Leung}}, \bibinfo
  {author} {\bibfnamefont {D.}~\bibnamefont {Zhu}}, \bibinfo {author}
  {\bibfnamefont {N.~M.}\ \bibnamefont {Linke}}, \bibinfo {author}
  {\bibfnamefont {K.~R.}\ \bibnamefont {Brown}}, \bibinfo {author}
  {\bibfnamefont {L.}~\bibnamefont {Duan}}, \ and\ \bibinfo {author}
  {\bibfnamefont {C.}~\bibnamefont {Monroe}},\ }\href@noop {} {\bibfield
  {journal} {\bibinfo  {journal} {Physical Review A}\ }\textbf {\bibinfo
  {volume} {100}},\ \bibinfo {pages} {022332} (\bibinfo {year}
  {2019})}\BibitemShut {NoStop}%
\bibitem [{\citenamefont {Milne}\ \emph {et~al.}(2020)\citenamefont {Milne},
  \citenamefont {Edmunds}, \citenamefont {Hempel}, \citenamefont {Roy},
  \citenamefont {Mavadia},\ and\ \citenamefont {Biercuk}}]{milne2020phase}%
  \BibitemOpen
  \bibfield  {author} {\bibinfo {author} {\bibfnamefont {A.~R.}\ \bibnamefont
  {Milne}}, \bibinfo {author} {\bibfnamefont {C.~L.}\ \bibnamefont {Edmunds}},
  \bibinfo {author} {\bibfnamefont {C.}~\bibnamefont {Hempel}}, \bibinfo
  {author} {\bibfnamefont {F.}~\bibnamefont {Roy}}, \bibinfo {author}
  {\bibfnamefont {S.}~\bibnamefont {Mavadia}}, \ and\ \bibinfo {author}
  {\bibfnamefont {M.~J.}\ \bibnamefont {Biercuk}},\ }\href@noop {} {\bibfield
  {journal} {\bibinfo  {journal} {Physical Review Applied}\ }\textbf {\bibinfo
  {volume} {13}},\ \bibinfo {pages} {024022} (\bibinfo {year}
  {2020})}\BibitemShut {NoStop}%
\bibitem [{\citenamefont {Bentley}\ \emph {et~al.}(2020)\citenamefont
  {Bentley}, \citenamefont {Ball}, \citenamefont {Biercuk}, \citenamefont
  {Carvalho}, \citenamefont {Hush},\ and\ \citenamefont
  {Slatyer}}]{bentley2020numeric}%
  \BibitemOpen
  \bibfield  {author} {\bibinfo {author} {\bibfnamefont {C.~D.}\ \bibnamefont
  {Bentley}}, \bibinfo {author} {\bibfnamefont {H.}~\bibnamefont {Ball}},
  \bibinfo {author} {\bibfnamefont {M.~J.}\ \bibnamefont {Biercuk}}, \bibinfo
  {author} {\bibfnamefont {A.~R.}\ \bibnamefont {Carvalho}}, \bibinfo {author}
  {\bibfnamefont {M.~R.}\ \bibnamefont {Hush}}, \ and\ \bibinfo {author}
  {\bibfnamefont {H.~J.}\ \bibnamefont {Slatyer}},\ }\href@noop {} {\bibfield
  {journal} {\bibinfo  {journal} {Advanced Quantum Technologies}\ }\textbf
  {\bibinfo {volume} {3}},\ \bibinfo {pages} {2000044} (\bibinfo {year}
  {2020})}\BibitemShut {NoStop}%
\bibitem [{\citenamefont {Green}\ and\ \citenamefont
  {Biercuk}(2015)}]{green2015phase}%
  \BibitemOpen
  \bibfield  {author} {\bibinfo {author} {\bibfnamefont {T.~J.}\ \bibnamefont
  {Green}}\ and\ \bibinfo {author} {\bibfnamefont {M.~J.}\ \bibnamefont
  {Biercuk}},\ }\href@noop {} {\bibfield  {journal} {\bibinfo  {journal}
  {Physical Review Letters}\ }\textbf {\bibinfo {volume} {114}},\ \bibinfo
  {pages} {120502} (\bibinfo {year} {2015})}\BibitemShut {NoStop}%
\bibitem [{\citenamefont {James}(1998)}]{james1998quantum}%
  \BibitemOpen
  \bibfield  {author} {\bibinfo {author} {\bibfnamefont {D.}~\bibnamefont
  {James}},\ }\href@noop {} {\bibfield  {journal} {\bibinfo  {journal} {Applied
  Physics B}\ }\textbf {\bibinfo {volume} {2}},\ \bibinfo {pages} {181}
  (\bibinfo {year} {1998})}\BibitemShut {NoStop}%
\bibitem [{\citenamefont {Wang}(2022)}]{patent2022wang}%
  \BibitemOpen
  \bibfield  {author} {\bibinfo {author} {\bibfnamefont {J.-B.}\ \bibnamefont
  {Wang}},\ }\href
  {https://patents.google.com/patent/CN114707358A/en?oq=CN114707358A} {\enquote
  {\bibinfo {title} {Ion trap quantum gate fidelity optimization method and
  device, electronic equipment and medium},}\ } (\bibinfo {year}
  {2022})\BibitemShut {NoStop}%
\end{thebibliography}%
\end{document}